\documentclass[DIV=calc, paper=letter, fontsize=11pt]{scrartcl} 
\usepackage[]{units}
\usepackage{lipsum} % Used for inserting dummy 'Lorem ipsum' text into the template
\usepackage[english]{babel} % English language/hyphenation
\usepackage[protrusion=true,expansion=true]{microtype} % Better typography
\usepackage{amsmath,amssymb,amsfonts,amsthm} % Math packages
\usepackage[svgnames, table]{xcolor} % Enabling colors by their 'svgnames'
\usepackage[hang, small,labelfont=bf,up,textfont=it,up]{caption} 
\usepackage{booktabs} % Horizontal rules in tables
\usepackage{fix-cm}	 
\usepackage[]{units}

\usepackage{tabularx} %used for table
\usepackage{multirow}
\usepackage{rotating}
\usepackage{array,makecell,multirow}

\usepackage[colorlinks=true, linkcolor=blue, citecolor=blue, 
filecolor=blue, runcolor=blue, urlcolor = blue]{hyperref}

\usepackage{fullpage}
\usepackage{sectsty} % Enables custom section titles

\usepackage{fancyhdr} % Needed to define custom headers/footers
\usepackage{lastpage} % Used to determine the number of pages 
\usepackage{nicefrac}
\usepackage{cite}

\usepackage{tikz}
\usetikzlibrary{calc,patterns,decorations.pathmorphing,decorations.markings}
\usepackage{pgfplots}

\usepackage{multicol}
\usepackage{float} % PJA: Note that need to avoid floating figures with multicols

\newcounter{tempcolnum}
\makeatletter
\newcommand{\multicolinterrupt}[1]{% Stuff to span both rows
\setcounter{tempcolnum}{\col@number}
\end{multicols}
#1%
\begin{multicols}{\value{tempcolnum}}
}
\makeatother

\newcommand{\atzEquNumAndAdd}{
(\arabic{equation}) 
\addtocounter{equation}{1}
}

\usepackage{graphicx}
\DeclareGraphicsExtensions{.png}

\usepackage{lettrine} % Package to accentuate the first letter of the text
\newcommand{\initial}[1]{ % Defines the command and style for the first letter
\lettrine[lines=3,lhang=0.3,nindent=0em]{
\color{DarkGoldenrod}
{\textsf{#1}}}{}}

\usepackage{caption}

\usepackage{float} % PJA: Note that need to avoid floating figures with multicols

\usepackage{graphicx}

\usepackage{titling} % Allows custom title configuration

\newcommand{\HorRule}{\color{DarkGoldenrod} 
\rule{\linewidth}{1pt}} % Defines the gold horizontal rule around the title

\pretitle{\vspace{-80pt} \begin{flushleft} 
\HorRule \fontsize{14.8}{14.8} \color{DarkRed} 
\selectfont} % Horizontal rule before the title

\title{Surface Fluctuating Hydrodynamics Methods for the Drift-Diffusion \\ 
Dynamics of Particles and Microstructures within Curved Fluid Interfaces}

\posttitle{\par\end{flushleft}} % Whitespace under the title 
\preauthor{\vspace{-10pt} \begin{flushleft}\fontsize{14}{14} 
\large  \color{DarkRed}} % Author font configuration
\author{David A. Rower$^{*}$, Misha Padidar$^{*}$, 
and Paul J. Atzberger$^{*}$ } % Your name
\postauthor{\fontsize{12}{12} \small \color{Black} 
\\ $*$ Department of Mathematics, University of California Santa Barbara; 
e-mail: atzberg@gmail.com; website: http://atzberger.org/;  \\
\vspace{0.5em}
\lfoot{\footnotesize * Work supported by DOE Grant ASCR PHILMS DE-SC0019246 
and NSF Grant DMS - 1616353.}% Your institution
\end{flushleft} \vspace{-18pt} \HorRule} % Horizontal rule after the title
\date{}

\newcommand{\mb}[1]{\mathbf{#1}}
\newcommand{\bs}[1]{\boldsymbol{#1}}

\newcommand{\ra}[1]{\renewcommand{\arraystretch}{#1}}

\newcommand{\parz}[1]{\left(#1\right)}
\definecolor{issuePJA_color}{rgb}{1.0,0.0,0.0}

\definecolor{commentPJA_color}{rgb}{1.0,0.0,0.8}

\definecolor{commentMP_color}{rgb}{1.0,0.5,0.0}

\DeclareGraphicsExtensions{.png}

\begin{document}

\maketitle % Print the title

\thispagestyle{fancy} % Enabling the custom headers/footers for the first page 

%------------------------------------------------------------------------------
%	ABSTRACT
%------------------------------------------------------------------------------
\vspace{-13pt}
% The first character should be within \initial{}
\initial{W}\textbf{e introduce fluctuating hydrodynamics approaches on surfaces
for capturing the drift-diffusion dynamics of particles and microstructures
immersed within curved fluid interfaces of spherical shape.  We take into
account the interfacial hydrodynamic coupling, traction coupling with the
surrounding bulk fluid, and thermal fluctuations.  For fluid-structure
interactions, we introduce Immersed Boundary Methods (IBM) and related
Stochastic Eulerian-Lagrangian Methods (SELM) for curved surfaces.  We use
these approaches to investigate the statistics of surface fluctuating
hydrodynamics and microstructures.  For velocity autocorrelations, we find
characteristic power-law scalings $\tau^{-1}$, $\tau^{-2}$, and plateaus can
emerge.  This depends on the physical regime associated with the geometry, surface
viscosity, and bulk viscosity.  This differs from the characteristic
$\tau^{-3/2}$ scaling for bulk three dimensional fluids.  We develop theory
explaining these observed power-laws associated with time-scales for 
dissipation within the fluid interface and coupling to the surrounding
fluid.  We then use our introduced methods to investigate a few example systems
and roles of hydrodynamic coupling and thermal fluctuations
including for the kinetics of passive particles and active microswimmers
in curved fluid interfaces.}

%------------------------------------------------------------------------------
%	ARTICLE CONTENTS
%------------------------------------------------------------------------------

\begin{multicols}{2}

\section{Introduction}
\label{sec:introduction} 
Soft materials can exhibit
rich mechanical responses arising 
from curved fluid interfaces that
mediate interactions between
immersed particles and other 
microstructures~\cite{Fuller2012,HonerkampSmith2013,Bassereau2011}.  
This includes protein interactions within lipid bilayer 
membranes~\cite{Alberts2007,HonerkampSmith2013,
AtzbergerBassereau2014,Ando2010},
surfactants and contaminants in bubble
interfaces~\cite{LealFeng1997,LealStone1990}, transport in soap
films~\cite{Kellay2017,Kornek2010}, and recent systems with nanoparticles or
colloids embedded in fluid interfaces~\cite{StebeCurvatureRodAssemblyPNAS2011,
Lee2016,AbbottColloidDroplets2014,Dominguez2018,
  Vlahovska2016,Choi2011,Bresme2007,DesernoMuellerGuven2005}.  Related
hydrodynamic and curvature mediated phenomena also play an important role in
biology and physiology, including transport of surfactants in lung
alveoli~\cite{Hermans2015,SquiresManikantan2017} or in cell
  mechanics~\cite{Mogilner2018,Chou2010,Powers2002}.  We develop general
approaches to model and simulate the collective drift-diffusion dynamics of
particles and other microstructures embedded within curved two-dimensional
fluid interfaces.  At small length and time scales, our methods 
also allow for capturing the roles of fluctuations arising from active
microstructures and from 
thermal effects~\cite{Saintillan2018,Reichl1997,Bird1987,Bresme2007,Cai1995}.

We develop methods for curved surfaces building on our prior work on
stochastic immersed boundary methods and fluctuating hydrodynamics
methods~\cite{AtzbergerSIB2007,AtzbergerSELM2011,AtzbergerSoftMatter2016}.  We
also draw on our recent work on developing approaches for deterministic
incompressible hydrodynamic flows on curved
surfaces~\cite{AtzbergerGrossHydro2018,GrossAtzbergerGMLSHydro2019}.  
Here, we
address how to introduce the spontaneous thermal fluctuations and handle the
associated drift-diffusive dynamics of microstructures both
for hydrodynamics in the inertial
regime and the overdamped quasi-steady regime.
We develop theory
and computational methods that capture for the fluctuations the correlations
arising from the hydrodynamic coupling within the curved fluid interface and
from the traction stresses with the flows of surrounding bulk fluids.

Many past approaches for investigating hydrodynamic
coupling and diffusion have been based on the classic Saffman-Delbr\"{u}ck (SD)
hydrodynamics model~\cite{SaffmanDelbruck1975,Saffman1976}.  
These were derived for
flat viscous sheets~\cite{Stone1998,Levine2004,
Oppenheimer2009,CamleyBrown2014,Demery2010,
Oppenheimer2017}.  
However, for many
problems arising in practice, the geometry plays an important role as a
consequence of significant curvature on the SD length-scale or from the surface
topology.  These effects can significantly change the hydrodynamic responses
relative to the flat case~\cite{AtzbergerSoftMatter2016,Rangamani2013,
  AtzbergerGrossHydro2018,HonerkampSmith2013}.  This has motivated recent work
going beyond the classic Saffman-Delbr\"{u}ck theory to take into account the
role of geometry and additional mechanical effects arising in curved fluid
interfaces~\cite{Henle2010,Arroyo2009,Rangamani2013,SahuSauerMandadapu2017,
Dominguez2018,AtzbergerSoftMatter2016, AtzbergerGrossHydro2018}.

We introduce general fluctuating hydrodynamics methods for the drift-diffusion
of particles and microstructures immersed within curved fluid interfaces.  To
demonstrate ideas, we focus particularly on the case of interfaces of spherical
shape.  To ensure finite diffusivities, as indicated in the SD theory, we 
formulate hydrodynamic equations for curved fluid interfaces 
coupled to surrounding bulk fluids.
We introduce thermal fluctuations accounting for the
fluid-structure interactions and collective dynamics of the particles and
microstructures building on our prior work on stochastic immersed boundary methods
and related approaches in~\cite{AtzbergerSIB2007, AtzbergerSELM2011,
AtzbergerSoftMatter2016,AtzbergerGrossHydro2018,GrossAtzbergerGMLSHydro2019}.

For curved fluid interfaces, the geometry and topology pose 
additional challenges for developing stochastic immersed boundary 
methods and numerical approaches.  This includes formulating 
and developing methods to solve hydrodynamic equations within 
curved surfaces and to obtain appropriate fluctuations and fluid-structure coupling 
operators~\cite{AtzbergerSoftMatter2016}.  We develop stochastic numerical methods
and approaches for fluid-structure coupling operators taking the geometry into 
account.  These numerical solvers and coupling operators allow us to 
obtain mobility tensors for the collective hydrodynamic coupling for modeling
both passive and active microstructures within the interface.

The geometry and topology of curved interfaces also pose additional
challenges for generating incompressible hydrodynamic
fields and thermal fluctuations.  We develop techniques to handle these
aspects of the fluid mechanics based on formulating generalized 
vector potentials using the Hodge decomposition for manifolds.  
For the overdamped quasi-steady regime,
we further develop techniques for the mobility tensor by using the 
embedding space to derive covariance structures that capture consistently 
hydrodynamic correlations in fluctuating fields.  The embedding space
expands the dimension of the problem and can create null-spaces for
some of the linear operators.
To address this issue, we develop algorithms based on stabilizations
for computing the stochastic driving forces in the drift-diffusion 
dynamics of the microstructures.
Our introduced approaches provide methods for both inertial and
overdamped quasi-steady regimes.  Our methods provide approaches
for capturing in simulations the interface hydrodynamics, 
traction coupling with the surrounding bulk fluid, 
fluid-structure coupling, and thermal fluctuations. 

We organize our paper as follows.  We formulate the fluctuating hydrodynamic
equations and related immersed boundary methods for spherical fluid interfaces
in Section~\ref{sec:fluct_hydro_curved}.  Stochastic numerical
methods for the drift-diffusion dynamics of microstructures are discussed in
Section~\ref{sec:stoch_methods}.  We use our approaches to investigate the
statistical mechanics of surface fluctuating hydrodynamics and to develop theory
for explaining observed power-laws in Section~\ref{sec:applications}.  We
further demonstrate our approaches by investigating the role of hydrodynamic
coupling and related diffusive correlations in the kinetics of passive
particles and active microswimmers in Section~\ref{sec:applications}. 
The
results show some of the rich phenomena that can arise for microstructures
and fluctuating hydrodynamics within curved fluid interfaces.

%%%%%%%%%%%%%%%%%%%%%%%%%%%%%%%%%%%%%%%%%%%%%
\lfoot{} % PJA removes comment about support from left footer after 
% first page (so does not appear on later pages).
% PJA: location of this empty lfoot can be sensitive which pages and later 
% it effects... we might be doing a LaTeX hack, so be careful of 
% placement of this empty.
%%%%%%%%%%%%%%%%%%%%%%%%%%%%%%%%%%%%%%%%%%%%%

\section{Fluctuating Hydrodynamics for Curved Fluid Interfaces}
\label{sec:fluct_hydro_curved}

To formulate the hydrodynamics of curved fluid interfaces, we will 
first develop the conservation laws with reference to the ambient 
embedding space.  We will then develop descriptions in terms 
of operators that generalize the techniques of vector 
calculus used in continuum 
mechanics~\cite{Marsden1994,AtzbergerGrossHydro2018}.

\subsection{Conservation Laws for Curved Surfaces}
\label{sec:conservation_laws}
The conservation of mass and momentum of the fluid can be expressed as  
\begin{eqnarray}
\label{equ_cont_mech_fluid}
\nonumber
\atzEquNumAndAdd \\
\nonumber
\left\{
\begin{array}{llll}
\rho \dot{\mb{v}} & = & \overline{\mbox{div}}\left(\bs{\sigma}\right) + \rho\bar{\mb{b}}\\
\dot{\rho} + \rho\left(\overline{\mbox{div}} 
\left(\mb{v}_{\parallel} \right) + \mb{v}_n H\right) & = & 0.
\end{array}
\right.
\end{eqnarray}
The $\mb{v}_n$ and $\mb{v}_\parallel$ denote respectively the components of the
fluid velocity normal and tangential to the fluid interface with 
$\mb{v} = \mb{v}_\parallel + \mb{v}_n$.  The $H$ denotes
the local mean curvature of the surface~\cite{Pressley2001}.  The $\rho$ is the
local mass density, $\bs{\sigma}$ the internal interfacial stress, and
$\overline{\mb{b}}$ the body force per unit mass.  

V
The $\overline{\mbox{div}} \left(\mb{t} \right) = \mb{t}_{|b}^b$ denotes the
surface covariant
divergence~\cite{Abraham1988,AtzbergerSoftMatter2016,AtzbergerGrossHydro2018}.
For a vector field represented as $\mb{t} = t^{a}\partial_{x^a}$ 
and $\mb{v} = v^a\partial_{x^a}$,
the surface covariant derivative $\nabla_{\mb{v}} \mb{t}$
gives
$\mb{w} = \nabla_{\mb{v}} \mb{t} 
= t_{|b}^c v^b \partial_{x^c} = w^c \partial_{x^c}$,
with $w^c = t_{|b}^c v^b$. 
The
$t_{|b}^c = {\partial t^c}/{\partial x^b} + \Gamma_{ab}^c t^a$ 
and
$\Gamma_{ab}^c$ denotes the Christoffel
symbols~\cite{Abraham1988,Pressley2001}.  
The material derivative of the mass on a surface
is $\dot{\rho} = \partial \rho/\partial{t} +
\rho\left(\overline{\mbox{div}}(\mb{v}_{\parallel}) + \mb{v}_n H\right)$,
where $H$ is the local mean curvature.  
The material derivative of the momentum 
is $\dot{\mb{v}} = \partial
\mb{v}/\partial t + \nabla_{\mb{v}}
\mb{v}$~\cite{Abraham1988,AtzbergerGrossHydro2018}.
In the manifold setting, the 
material derivative of the vector field
$\mb{t}$ can be expressed as
$\dot{\mb{t}} = L_{\mb{v}}\mb{t}$,
where $L_{\mb{v}}$ is the Lie derivative of 
$\mb{t}$ under the flow of the velocity field
$\mb{v}$~\cite{Abraham1988,Marsden1994}.

Throughout, we shall consider Newtonian incompressible fluid interfaces of
fixed shape.  In this case, the hydrodynamic flows are tangential to the
surface.  As a consequence, $\mb{v}_n = 0$, $\rho = \rho_0$, $\dot{\rho} = 0$
so that $\overline{\mbox{div}}\left(\mb{v}\right) = 0$.  For notational 
convenience, we suppress the $\mb{v}_{\parallel}$ and $\rho_0$ for the 
velocity and mass denoting them simply as $\mb{v}$ and $\rho$.  The
conservation laws simplify in this setting to
\begin{eqnarray}
\label{equ_cont_mech_fluid2} \left\{ \begin{array}{llll} \rho \dot{\mb{v}} &
= & \overline{\mbox{div}}\left(\bs{\sigma}\right) + \mb{b}\\
\overline{\mbox{div}}\left(\mb{v} \right) & = & 0.  \end{array} \right.
\end{eqnarray} 
The $\rho$ is now constant throughout and $\mb{b}$ denotes the local 
force density.
The stress tensor $\bs{\sigma}$ for an incompressible Newtonian
fluid for a curved fluid interface can be expressed as \begin{eqnarray}
\label{equ_sigma_fluid} \bs{\sigma} = \mu_m \mb{D} - p \mathcal{I}.
\end{eqnarray} 
The $\mu_m$ is the surface viscosity of the interfacial fluid,
$\mb{D}$ is the rate-of-deformation tensor, $p$ is the
pressure, and $\mathcal{I}$ is the metric associated identity
tensor~\cite{Marsden1994}.  

Tensors can be expressed in covariant or contravariant
form~\cite{Heinbockel2001,Abraham1988}.  For vectors in contravariant form we
have $\mb{t}^{\sharp} = t^a \partial_{x^a}$ or in covariant form
$\mb{t}^{\flat} = t_a \mb{d}x^a$~\cite{Abraham1988}.  The $\partial_{x^a}$
denotes the $a^{th}$ coordinate basis vector.  The $\mb{d}x^a$ denotes the
$a^{th}$ coordinate basis covector (differential $1$-form) with
$\mb{d}x^a[\partial_{x^b}] = \delta_a^b$~\cite{Abraham1988}.  We can convert
between vectors and covectors by the mappings $\flat: t^a \rightarrow t_a =
g_{ab}t^b$ and $\sharp: t_a \rightarrow t^a = g^{ab}t_b$.  The $g_{ab}$ denotes
the metric tensor and $g^{ab}$ the inverse metric tensor~\cite{Abraham1988}.
The isomorphic maps $\flat$ and $\sharp$ between the 
tangent space and cotangent space correspond operationally in calculations
to lowering and raising indices in the coordinate expressions of the
tensors~\cite{Heinbockel2001,Abraham1988}.

We find it convenient in our calculations to express tensors in covariant form
and use exterior calculus~\cite{Abraham1988}.  This allows us to generalize
vector calculus and many techniques employed for fluid mechanics to the
manifold
setting~\cite{AtzbergerGrossHydro2018,
GrossAtzbergerGMLSHydro2019,AtzbergerSoftMatter2016}.
For an incompressible Newtonian fluid interface, the stress $\bs{\sigma}$ is
given in equation~\ref{equ_sigma_fluid}. The divergence of the stress tensor on
the surface $\overline{\mbox{div}}\left(\bs{\sigma}\right)$ becomes in
covariant
form~\cite{AtzbergerGrossHydro2018,AtzbergerSoftMatter2016,Marsden1994}
\begin{eqnarray} \overline{\mbox{div}}\left(\bs{\sigma}\right)^{\flat} = \mu_m
\left(-\bs{\delta} \mb{d} \mb{v}^{\flat} + 2K \mb{v}^{\flat}\right) - \mb{d}p.
\end{eqnarray} 
The $K$ is the local
Gaussian curvature of the surface~\cite{Pressley2001}.
The $\mb{d}$ is the exterior derivative playing a role
similar to the gradient of the vector field on the surface.   For a 1-form 
$\mb{v}^{\flat} = v^i d\mb{x}_i$ the exterior derivative is given by 
$\mb{d}\mb{v}^{\flat} = \partial v^i/\partial x^j\, d\mb{x}_j \wedge d\mb{x}_i$.
The
$\bs{\delta}=-\star\mb{d}\star $ 
is the co-differential playing a role similar to the 
divergence on the surface~\cite{Abraham1988}.   The $\star$ is the Hodge star which
for a differential $k$-form $\bs{\beta}$ gives a complementary $n-k$-form 
$\star \bs{\beta}$ so that for any $k$-form $\bs{\alpha}$ we have
$\bs{\alpha}\wedge \star \bs{\beta} = \langle\alpha,\beta\rangle \bs{\omega}$
where $\bs{\omega}$ is the volume form~\cite{Abraham1988}.  This also allows
us to generalize vector calculus operators such as the curl and divergence
to the surface by $\mbox{curl}_\mathcal{M}(\mb{v}^{\flat})
= -\star \mb{d} \mb{v}^{\flat}$
and $\mbox{div}_\mathcal{M}(\mb{v}^{\flat}) = \bs{\delta}\mb{v}^{\flat}$.
We give coordinate
expressions for these operations and additional discussions 
in Appendix~\ref{appendix:exterior_calc_coord}

\subsection{Fluctuating Hydrodynamics with Fluid-Structure Interactions}
\label{sec:SELM_curved} 
\begin{figure}[H] \centering
\includegraphics[width=0.9\columnwidth]
{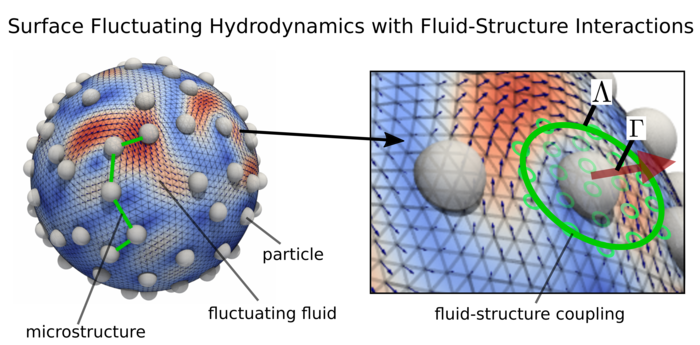}
\caption{Surface Fluctuating Hydrodynamics with Fluid-Structure Interactions.
To model particles and microstructures embedded in curved fluid interfaces, we
couple a Lagrangian description of the microstructures to an Eulerian
description for the fluid mechanics.  The coupling is modeled by the two
operators $\Gamma$ and $\Lambda$. The operator $\Gamma$ gives a kinematic
reference fluid velocity used to determine how the fluid exerts force upon
microstructures in equation~\ref{equ:fluct_hydro_microstructure}. The operator
$\Lambda$ gives the related force density for how the microstructures exert
force on the fluid in equation~\ref{equ:fluct_hydro_fluid}.  We model the
drift-diffusion dynamics of microstructures by the SELM fluctuating
hydrodynamics description introduced in equations~\ref{equ:fluct_hydro_fluid} -
\ref{equ:fluct_hydro_microstructure_V}.  } \label{fig:fluct_hydro_coupling}
\end{figure}

We develop surface fluctuating hydrodynamics descriptions to
account for the drift-diffusion motions of microstructures and their
hydrodynamic coupling within curved 
fluid interfaces~\cite{AtzbergerSIB2007,AtzbergerSELM2011,
AtzbergerSoftMatter2016}.  In covariant form, we introduce for curved fluid
interfaces fluctuating hydrodynamic equations incorporating fluid-structure
interactions.   We illustrate our general approach in
Figure~\ref{fig:fluct_hydro_coupling}.   

The fluid dynamics are modeled by 
\begin{eqnarray}
\label{equ:fluct_hydro_fluid}
\rho \frac{d\mb{v}^{\flat}}{dt} & = & 
\mu_m 
\left(
-\bs{\delta} \mb{d} \mb{v}^{\flat}
+ 
2K \mb{v}^{\flat}
\right)
- \mb{d}p \\
\nonumber
& + & 
\mb{t}^{\flat}
\\
\nonumber
& + & 
 \Lambda\left[
 \gamma\left(
\mb{V} - \Gamma \mb{v}^{\flat}
\right)
\right]
+ \mb{f}_{thm}^{\flat} \\
-\bs{\delta} \mb{v}^{\flat} & = & 0.
\end{eqnarray}
The drift-diffusion motions of microstructures are modeled by
\begin{eqnarray}
\label{equ:fluct_hydro_microstructure}
m\frac{d\mb{V}}{dt} & = & -\gamma\left(\mb{V} - \Gamma \mb{v}^{\flat}\right) 
-\nabla \phi + \mb{F}_{thm} \\
\label{equ:fluct_hydro_microstructure_V}
\frac{d\mb{X}}{dt} & = & \mb{V}.
\end{eqnarray}
The $\mb{X}$ and $\mb{V}$
denotes the collective configuration and velocity of the microstructures
immersed within the fluid. The $\mb{v}^{\flat}$ denotes the fluid velocity
and the
$-\bs{\delta} \mb{v}^{\flat} = 0$ enforces the local incompressibility 
of the flow.
The $\mu_m$ is the surface fluid shear viscosity, 
$p$ is the pressure, $K$ is the local Gaussian curvature, $\bs{t}^{\flat}$
is the traction stress with the surrounding fluid.
The $-\nabla{\phi}$ are the conservative
forces acting on the microstructures, and $\mb{f}_{thm}$ and $\mb{F}_{thm}$ are
the stochastic forces accounting for thermal fluctuations of the system.

V
The fluid-structure interactions result in two equal-and-opposite forces.
The
$
-\gamma\left(
\mb{V} - \Gamma \mb{v}^{\flat}
\right)
$
is a drag force of strength $\gamma$ the 
microstructures experience from local coupling to the fluid.
The 
$ 
 \Lambda\left[
 \gamma\left(
\mb{V} - \Gamma \mb{v}^{\flat}
\right)
\right]
$ is the opposite force density the fluid experiences 
locally from coupling to the microstructures.
The $\Lambda$ is a spreading operator 
that serves to convert a local force to a local force density acting on
the fluid.  The $\Gamma$ is an averaging operator that serves to estimate a 
local reference velocity for a microstructure from the nearby fluid flow.
We discuss specific choices for $\Lambda$ and
$\Gamma$ in more detail
in Section~\ref{sec:IB_curved}.  
V
Given the rapid oscillations of the fluid from the thermal 
fluctuations, we neglect the advection terms
which are expected to give lower-order contributions in
equation~\ref{equ:fluct_hydro_fluid}~\cite{AtzbergerTabak2015}.

\end{multicols} 

\begin{multicols}{2} \noindent

\subsection{Traction Stress from Flow of the Surrounding Bulk Fluid}
To obtain the traction stresses
$\mb{t}^{\flat} = \mathcal{T}_f \mb{v}^{\flat}$ for the 
spherical geometry, we use Lamb's
solution~\cite{HappelBrenner1983,Lamb1895}. The traction arises from 
the surface flow with velocity $\mb{v}$ entraining the surrounding 
bulk fluid~\cite{AtzbergerSoftMatter2016}.
Our approach makes the assumption that the bulk surrounding flow arises
from an incompressible Newtonian fluid that reaches steady-state rapidly when
contributing to the surface traction.  Let the bulk fluid velocity be denoted by
$\mb{u}$ with values on the surface $\mb{u} = \mb{v} + v_n \mb{n}$, where we
shall assume $v_n = 0$ throughout.  The solution can be expressed using the
spherical harmonics expansion 
\begin{eqnarray} 
\label{equ_Z_n} \mb{r}\cdot\nabla \times \mb{u} 
& = & \sum_{\ell = -\infty}^{\infty} Z_\ell.
\end{eqnarray} We emphasize here the curl $\nabla \times$ is the usual operator
in three dimensional Euclidean space.  The $Z_\ell$ denotes the combined
contributions of all of the solid spherical harmonic terms of degree $\ell$.
We also expand the bulk surrounding fluid flows inside the sphere $\mb{u}^{-}$
and outside the sphere $\mb{u}^{+}$ as 
\begin{eqnarray} 
\mb{u}^{+} = \sum_{\ell= 0}^{\infty}
\mb{u}^{+}_\ell, \hspace{0.4cm} \mb{u}^{-} 
= \sum_{\ell = 1}^{\infty} \mb{u}^{-}_\ell.  
\end{eqnarray} 
The $\mb{u}_\ell^{+}$ and
$\mb{u}_\ell^{-}$ are the solid spherical harmonic expansion terms combined for
a given degree $\ell$.  
The Lamb solutions for the bulk fluid velocity fields are given
by~\cite{HappelBrenner1983,Lamb1895} expressible as
\begin{eqnarray} 
\label{equ_lambs_sol}
\mb{u}^{+}_\ell 
= \nabla \times \left(\mb{r}\chi_{-(\ell+1)}\right),
\hspace{0.4cm} \mb{u}^{-}_\ell 
= \nabla \times \left(\mb{r}\chi_\ell\right),
\end{eqnarray}
where \begin{eqnarray} \chi_\ell & = & \frac{1}{\ell(\ell + 1)}
\left(\frac{r}{R}\right)^\ell Z_\ell \\ \chi_{-(\ell+1)} & = &
\frac{1}{\ell(\ell + 1)} \left(\frac{R}{r}\right)^{\ell+1} Z_{\ell}.
\end{eqnarray}
The traction stress expressed in contravariant form $\mb{t}^{\sharp} =
\mb{t}^{+} + \mb{t}^{-}$ is 
\begin{eqnarray} 
\label{equ_tract_stress_tplus}
\mb{t}^{+} = \bs{\sigma}^{+}\cdot \mb{n}^{+} = \mu_{+} 
\frac{\partial \mb{u}^{+}}{\partial r} 
+ \mu_{+} \nabla \left(\mb{u}^{+}\cdot \mb{n}^{+}\right) 
\hspace{0.4cm}\\ 
\label{equ_tract_stress_tminus} 
\mb{t}^{-} = \bs{\sigma}^{-}\cdot \mb{n}^{-} 
= -\mu_{-} \frac{\partial \mb{u}^{-}}{\partial r} 
+ \mu_{-} \nabla \left(\mb{u}^{-}\cdot \mb{n}^{-}\right).  
\end{eqnarray}
 
The $\bs{\sigma}^{\pm}$ is the stress of the 
bulk surrounding fluid arising at the fluid interface.  
For the traction stress 
$\mb{t}^{\pm} = \mb{t}^{\pm}(\mb{x}_0)$ on the surface,
note the $\mb{n}^{\pm}=\mb{n}^{\pm}(\mb{x}_0)$ is evaluated 
at the fixed location $\mb{x}_0$ while $\mb{u}^{\pm} = \mb{u}^{\pm}(\mb{x})$.  
When taking the 
gradient the $\mb{n}^{\pm}$ does not change, so that 
$\nabla (\mb{u}^{\pm}\cdot\mb{n}^{\pm}) = \nabla \mb{u}^{\pm} \cdot \mb{n}^{\pm}$.  
We take $\mu_{\pm} =
\mu_f$ throughout.    The traction can be expressed in terms of the spherical
harmonics expansion as \begin{eqnarray} \label{equ_tract_stress_u_p} \mb{t}^{+}
& = & -\mu_{+}\sum_{\ell = 0}^{\infty} \frac{(\ell + 2)}{R} \mb{u}_\ell^{+} =:
\tilde{\mathcal{T}}_f^{+} \mb{v}^{\sharp} \\ \label{equ_tract_stress_u_m}
\mb{t}^{-} & = & -\mu_{-}\sum_{\ell = 1}^{\infty} \frac{(\ell - 1)}{R}
\mb{u}_\ell^{-}  =: \tilde{\mathcal{T}}_f^{-} \mb{v}^{\sharp}.  \end{eqnarray}
The bulk fluid flow has $\mb{u}^{+} = \mb{u}^{-} = \mb{v}$ on the fluid
interface and is completely determined by $\mb{v}$.  In covariant
form, we define the traction operators on the surface corresponding to
equation~\ref{equ_tract_stress_u_p} and~\ref{equ_tract_stress_u_m} as
$\mb{t}^{\flat} = \mathcal{T}_f \mb{v}^{\flat} = \mathcal{T}_f^{+}
\mb{v}^{\flat} + \mathcal{T}_f^{-}\mb{v}^{\flat}$.

We remark that the approaches we develop can also be 
extended readily for surrounding fluids in systems that 
are subject to external shear flows or 
time-dependent flow responses.  This would 
correspond to $\mb{t}^{\flat}$ 
obtained from 
equations~\ref{equ_tract_stress_tplus}-\ref{equ_tract_stress_tminus} 
being computed from stresses of such flows from 
other model hydrodynamic equations or results from 
numerical solvers~\cite{Delgado_Buscalioni_2018,
HonerkampSmith2013,Woodhouse2012,Dominguez2018,AtzbergerSELM2011}.  
For such surrounding fluids, with flows having time-dependence
or flows having non-Newtonian responses, our approaches can
also be used capture these dissipitative contributions to 
the mechanics and develop the associated thermal 
fluctuations~\cite{AtzbergerSELM2011,
AtzbergerTabak2015}.

\subsection{Thermal Fluctuations} To determine the associated stochastic
driving terms $\mb{f}_{thm}(\mb{x},t)$ and $\mb{F}_{thm}(t)$ that account for
thermal fluctuations, we use an approach related to our Stochastic Eulerian
Lagrangian Method (SELM) framework~\cite{AtzbergerSELM2011,AtzbergerTabak2015}.
This gives $\mb{f}_{thm}(\mb{x},t)$ and $\mb{F}_{thm}(t)$ as
Gaussian processes with $\delta$-correlation in time, mean
zero, and covariances 
\begin{eqnarray} 
\label{equ_stoch_driving_fluid}
\left\langle \mb{f}_{thm}(t) \mb{f}_{thm}(s)^T \right \rangle & = & -2k_B{T}
\mathcal{L}_{ff}\delta(t - s) \hspace{1cm}\\ 
\label{equ_stoch_driving_struct}
\left\langle \mb{F}_{thm}(t) \mb{F}_{thm}(s)^T \right \rangle & = &
-2k_B{T}\mathcal{L}_{pp} \delta(t - s)\\ 
\label{equ_stoch_driving_cross}
\left\langle \mb{F}_{thm}(t) \mb{f}_{thm}(s)^T \right \rangle & = &
  2k_B{T}\mathcal{L}_{pf} \delta(t - s).  
\end{eqnarray} 

The notation is to be interpreted 
as taking an expectation, $\langle Z \rangle = \mathbb{E}[Z]$.  
The equation \ref{equ_stoch_driving_fluid} involves the 
dissipative operator $\mathcal{L}_{ff}\mb{v}^{\flat}$ associated
with the fluid and is defined by 
\begin{eqnarray}
\label{equ:Lff_operator} \mathcal{L}_{ff} & = & \mathcal{L}_f 
+ \mathcal{L}_c,\hspace{0.4cm} \mathcal{L}_c = - \gamma \Lambda
\Gamma, \\ 
\label{equ:Lf_operator} \mathcal{L}_{f} & = & \mu_m
\left(-\bs{\delta}\mb{d} + 2K\right) + \mathcal{T}_f.  
\end{eqnarray}
The terms in equation \ref{equ_stoch_driving_struct} arise from the 
dissipative operator $\mathcal{L}_{pp}$ of the microstructure degrees of
freedom and is defined by
$\mathcal{L}_{pp} \mb{V} := -\gamma \mathcal{I} \mb{V}$.  
The terms in equation~\ref{equ_stoch_driving_cross}
gives the cross-correlation that arises from microstructure-fluid
coupling giving the operator $\mathcal{L}_{pf}$ and is defined by 
$\mathcal{L}_{pf} \mb{v}^{\flat} := \gamma \Gamma \mb{v}^{\flat}$.
%the coupling terms
%$\mathcal{L}_{fp} \mb{V} := \gamma \Lambda \mb{V}$ and}

To express and generate efficiently the thermal fluctuations, 
it is useful to define a term $\mb{g}_{thm}$.  This term is taken to 
be independent of $\mb{F}_{thm}$ and to have 
covariance 
\begin{eqnarray} \label{equ_stoch_driving_fluid_gthm} 
\left\langle
\mb{g}_{thm}(t) \mb{g}_{thm}(s)^T \right \rangle & = & -2k_B{T}
\mathcal{L}_{f}\delta(t - s). \hspace{1cm} 
\end{eqnarray} 
We can then express the thermal fluctuations for 
the hydrodynamics as $\mb{f}_{thm} = \mb{g}_{thm}
- \Lambda[\mb{F}_{thm}]$. This gives 
the correlations in equations~\ref{equ_stoch_driving_fluid}
--~\ref{equ_stoch_driving_cross}.

In our formulation, we use throughout that
the fluid-structure coupling operators are adjoints
$\Lambda = \Gamma^T$, in the sense $\int_{\mathcal{S}} \langle \Lambda \mb{U},
\mb{u}^{\flat}\rangle_g dA = \langle \mb{U}, \Gamma \mb{u}^{\flat}\rangle$ for
all choices of $\mb{U}$ and
$\mb{u}^{\flat}$~\cite{AtzbergerSoftMatter2016,AtzbergerSELM2011}.  We discuss
choices for the coupling operators and the adjoint conditions in more detail in
Section~\ref{sec:IB_curved}.  For the stochastic driving terms of
equations~\ref{equ:fluct_hydro_fluid}--~\ref{equ:fluct_hydro_microstructure_V}
and~\ref{equ_stoch_driving_fluid}--~\ref{equ_stoch_driving_cross}, our SELM
approach ensures the Gibbs-Boltzmann distribution is 
invariant under the stochastic dynamics and
satisfies
detailed-balance~\cite{AtzbergerSELM2011,AtzbergerTabak2015,Reichl1998}.  
The stochastic
equations~\ref{equ:fluct_hydro_fluid}--~\ref{equ:fluct_hydro_microstructure_V}
should be interpreted in the sense of Ito
Calculus~\cite{Oksendal2000,Gardiner1985}.\\

We remark that the covariance operators in
equations~\ref{equ_stoch_driving_fluid}--\ref{equ_stoch_driving_cross} are to
be interpreted in the weak
sense~\cite{Lieb2001,AtzbergerRD2010,AtzbergerSELM2011}.  Consider
$\mb{u}(\mb{x}) = \int \int \bs{\alpha}(\mb{z},r,\mb{x})^T
\mb{f}_{thm}(\mb{z},r) d\mb{z} dr$ and $\tilde{\mb{u}}(\mb{y}) = \int\int
\tilde{\bs{\alpha}}(\mb{w},q,\mb{y})^T \mb{f}_{thm}(\mb{w},q) d\mb{w} dq$  and
$\mb{U} = \int \mb{A}(r)^T \mb{F}_{thm}(r) dr$ and $\tilde{\mb{U}} = \int
\tilde{\mb{A}}(q)^T \mb{F}_{thm}(q) dq$.  The $\bs{\alpha}$,
$\tilde{\bs{\alpha}}$, $\mb{A}$, $\tilde{\mb{A}}$ are smooth fields and vectors
playing the role of test functions.~\cite{Lieb2001}.  The associated
covariances are \begin{eqnarray} \nonumber \langle \mb{u}(\mb{x})
\tilde{\mb{u}}(\mb{y})^T \rangle =  \hspace{5cm} \atzEquNumAndAdd \\ \nonumber
-\int \int \int \int \bs{\alpha}(\mb{z},r,\mb{x})^T 2k_B{T} \mathcal{L}_{ff}
\tilde{\bs{\alpha}}(\mb{w},q,\mb{y}) d\mb{w}d\mb{z} \cdot \\ \nonumber \delta(r
- q) dr dq.  \end{eqnarray} The differential operator $\mathcal{L}_{ff}$
acts in the parameter $\mb{w}$.  We also have the covariances \begin{eqnarray}
\\ \nonumber \langle \mb{U} \tilde{\mb{U}}^T \rangle = \int \int 2k_B{T}\gamma
\mb{A}(r)^T\tilde{\mb{A}}(q) \delta(r - q) dr dq, \end{eqnarray} and
\begin{eqnarray} \nonumber \atzEquNumAndAdd \\ \nonumber \langle \mb{U}
\tilde{\mb{u}}(\mb{y})^T \rangle = -\int \int \int 2k_B{T}\gamma \mb{A}(r)^T
\Gamma \tilde{\bs{\alpha}}(\mb{w},q,\mb{y}) \cdot \\ \nonumber \delta(r - q)
d\mb{w} dr dq.  \end{eqnarray} The operator $\Gamma$ acts in the
parameter $\mb{w}$.  Additional discussions on how to interpret these operator
covariances also can be found in~\cite{AtzbergerRD2010,AtzbergerSELM2011}.

\subsection{Fluid-Structure Coupling: Immersed Boundary Methods for Curved
Surfaces}

\label{sec:IB_curved} We handle the fluid-structure interactions between the
microstructures with the fluid by developing extended immersed
boundary methods in the manifold
setting, see Figure~\ref{fig:fluct_hydro_coupling},
\cite{Peskin2002,AtzbergerSIB2007,AtzbergerSELM2011}.  
Many choices can be made for the
operators $\Gamma$ and $\Lambda$~\cite{AtzbergerSELM2011}.  To ensure that the
approximate fluid-structure coupling which converts between the Eulerian and
Lagrangian reference frames are non-dissipative, the operators are taken to be
adjoints~\cite{AtzbergerSELM2011,Peskin2002,AtzbergerTabak2015}.  

We require the coupling operators satisfy the following adjoint conditions for
any choice of test field $\mb{v}$ and vector $\mb{F}$,
\begin{eqnarray}
\label{equ_adjoint_cond} \langle 
\Gamma \mb{v} , \mb{F} \rangle & = & \langle
\mb{v} , \Lambda \mb{F} \rangle, 
\end{eqnarray} where the inner-products are
defined as 
\begin{eqnarray} 
\langle \Gamma \mb{v} , \mb{F} \rangle & = & \sum_i
\left[ \Gamma \mb{v} \right]_i \cdot \left[\mb{F}\right]_i \\ \langle \mb{v} ,
\Lambda \mb{F} \rangle & = & \int_{\Omega} \mb{v}(\mb{x})\cdot \left(\Lambda
\mb{F}\right)(\mb{x}) d\mb{x}.  
\end{eqnarray} 
The $\mb{X}$ denotes the
collective vector of particle locations, or more abstract microstructure
degrees of freedom~\cite{AtzbergerSELM2011}. 
For instance, the $i^{th}$ particle would be at location 
$\left[\mb{X}\right]_i$.  The $\cdot$ denotes the vector dot-product induced by
the ambient physical space.  For vector fields on the surface $\mb{v}$ and
$\mb{u}$ we have $\mb{v}(\mb{x}) \cdot \mb{u}(\mb{x}) =
\langle\mb{v}(\mb{x}),\mb{v}(\mb{x})\rangle_g = v^a g_{ab} u^b$.  We use the
notation $\Gamma^T = \Lambda$ to denote the adjoint
condition~\ref{equ_adjoint_cond}.  

In the setting of curved surfaces, the coupling operators $\Gamma$ and $\Lambda$
should be chosen carefully to control the velocity averaging and force spreading.
The $\Gamma$ should ensure the velocity averaging takes into account 
over the surface the different tangential directions at each location.
The $\Lambda$ for force densities should have 
well-controlled components for the surface in the tangential and 
normal directions.  We develop operators of the form
\begin{eqnarray} \Gamma \mb{v} & = & \int_{\Omega} \mb{W}\left[\mb{v}\right]
(\mb{y}) d\mb{y} \\ \Lambda \mb{F} & = & \mb{W}^*\left[ \mb{F} \right](\mb{x}).
\end{eqnarray} We use a tensor $\mb{W}$ to sample and weight values on the
surface to perform velocity averaging.  We use the adjoint tensor $\mb{W}^*$ to
produce a corresponding force density field on the surface compatible with our
adjoint conditions~\ref{equ_adjoint_cond}.  For the curved surface, we use the
geometrically motivated forms \begin{eqnarray} \mb{W}\left[\mb{v}\right] & = &
\sum_i  \mb{w}^{[i],\alpha}\left[\mb{v}\right]
\partial_{x^\alpha}|_{\mb{X}^{[i]}} \\ \mb{W}^*\left[\mb{F}\right] & = & \sum_i
\left(w^{[i],\alpha}\right)^{\gamma} F^{\alpha}      \partial_{x^\gamma}.
\end{eqnarray} 
The sum $i$ runs over the indices of the particle or microstructure and the
$\partial_{x^\alpha}|_{\mb{X}^{[i]}}$ denotes the tangent basis vector in
direction $\alpha$ at location $\mb{X}^{[i]}$.   We refer to the vector field
$\mb{w}^{[i],\alpha}$ as the probing vector field for direction $\alpha$.  

On the sphere, the rotational symmetry can be utilized.  For this case
with the spherical coordinates $(\theta,\phi)$, we take our probing vector
fields to be of the form $\mb{w}^{[i],\theta} = \psi(\mb{x} - \mb{X}^{[i]})
\partial_{\theta}$ and $\mb{w}^{[i],\phi} = \left(\psi(\mb{x} -
\mb{X}^{[i]})/\cos(\theta)\right) \partial_{\phi}$, where $\psi(r) = C
\exp(-r^2/2\sigma^2)$.  In practice, we truncate at length $r_0 = 4\sigma$ and
use $C$ to normalize so that $\psi(r)$ averages to one on the surface.

\subsection{Overdamped Limit} \label{sec:overdamped_limit}

In physical regimes with small Reynolds numbers $Re \ll 1$, we
consider the overdamped limit of the fluctuating hydrodynamic
equations~\ref{equ:fluct_hydro_fluid}--~\ref{equ:fluct_hydro_microstructure}
~\cite{AtzbergerSELM2011,AtzbergerTabak2015}.
In this regime, the limiting fluctuating hydrodynamic 
equations can be expressed as 
\begin{eqnarray} 
\label{equ_full_BD_model} 
&& \frac{d\mb{X}}{dt} = M
\mb{F} +  k_B{T} \nabla\cdot M + \mb{F}_{thm} \\ \nonumber && \langle
\mb{F}_{thm}(s) \mb{F}_{thm}(t)^T\rangle = 2k_B{T}M\delta(t - s),
\end{eqnarray} 
where 
\begin{eqnarray} \label{equ_Mobility_slip} 
M & = &
{\gamma^{-1}}\mathcal{I} + \Gamma \mathcal{S} \Lambda.  
\end{eqnarray}
This is obtained by taking the overdamped limit while retaining the 
finite slip term
$-\gamma\left(\mb{V} - \Gamma \mb{v}^{\flat} \right)$ in
equation~\ref{equ:fluct_hydro_fluid} and~\ref{equ:fluct_hydro_microstructure}.
This results in the slip term ${\gamma^{-1}}\mathcal{I}$ in
equation~\ref{equ_Mobility_slip}~\cite{AtzbergerTabak2015}.  In the
strong-coupling limit with $\gamma \rightarrow \infty$ the mobility simplifies to
\begin{eqnarray} \label{equ_Mobility_no_slip} M & = & \Gamma \mathcal{S}
\Lambda.  \end{eqnarray} The velocity averaging operator $\Gamma$ and force
spreading operator $\Lambda$ are as discussed in Section~\ref{sec:IB_curved}.
The $\mathcal{S}$ denotes the solution operator $\mb{v}^{\flat} =
\mathcal{S}\mb{b}^{\flat}$ for the hydrodynamic equations
\begin{eqnarray} \label{equ_steady_hydro} \left(\mu_m \left(-\bs{\delta}\mb{d}
+ 2K\right) + \mathcal{T}_f\right) \mb{v}^{\flat} = -\mb{b}^{\flat} \\
-\bs{\delta} \mb{v}^{\flat} = 0.  \end{eqnarray} The $\mathcal{T}_f$
gives the traction stresses with the surrounding bulk fluid as
discussed in Section~\ref{sec:SELM_curved}.  The $\mb{v}^{\flat} =
\mathcal{S}\Lambda \mb{F}$ is the solution in the case with $\mb{b} =
\Lambda\left[\mb{F}\right]$, where in practice we have $\mb{F} =
-\nabla \phi$.

To simply expressions, we also take for
equation~\ref{equ_full_BD_model} the limit of no-slip between the
microstructure and the fluid, which corresponds formally to $\gamma/m
\rightarrow \infty$.  This yields $d\mb{X}/dt = \mb{V}= \Gamma \mb{v}^{\flat}$
throughout.  Putting this together we arrive at the first term in
equation~\ref{equ_full_BD_model}.  For the full stochastic system, we perform
related detailed dimensional analysis and asymptotics
in~\cite{AtzbergerSELM2011,AtzbergerTabak2015,AtzbergerSoftMatter2016}.  
The equation~\ref{equ_full_BD_model} should be given the Ito
interpretation.  The thermal fluctuations involve configuration-dependent
correlations resulting in the spontaneous 
drift term $k_B{T} \nabla\cdot
\mb{M}$,~\cite{Oksendal2000,Gardiner1985,AtzbergerSELM2011,AtzbergerTabak2015}.
Additional analysis and reductions for different physical 
regimes can be found
in~\cite{AtzbergerSELM2011,AtzbergerTabak2015,AtzbergerSoftMatter2016}. 

\subsection{Formulating Fluctuating Hydrodynamics on Surfaces using Vector
Potentials $\Phi$} \label{sec:hydrodynamic_responses}
\label{sec:hydrodynamic_vec_poten_formulation} 
The hydrodynamic responses both in the inertial fluctuating hydrodynamics of
equation~\ref{equ:fluct_hydro_fluid} and in the overdamped regime of
equation~\ref{equ_steady_hydro} require computation of the operator
$\mathcal{L}_f$ of equation~\ref{equ:Lf_operator}.  Since the hydrodynamic
fields are incompressible, we can use the Hodge decomposition of the fluid
velocity $\mb{v}^{\flat} = \mb{d}\tilde{\Phi} + \bs{\delta} \Psi + h$.  The 
$\tilde{\Phi}$ and $\Psi$ are vector potentials and $h$ is a harmonic function
related to the topology of the manifold~\cite{Marsden1994,Abraham1988}.
The
$\Delta_H = \mb{d} \bs{\delta} + \bs{\delta} \mb{d} $ is the Hodge Laplacian and
the $h$ is harmonic in the sense $\Delta_H h = 0$.  For the case of
spherical geometry with tangential velocity fields, 
we can express incompressible flows as $\mb{v}^{\flat} =
-\star\mb{d}\Phi$,~\cite{AtzbergerSoftMatter2016}.  This uses that 
incompressibility requires $\bs{\delta} \mb{v}^{\flat} = 0$
and from the spherical geometry $h = 0$.
This simplifies the action
of the Hodge Laplacian and the representations.  This allows for
reformulating the inertial hydrodynamics of equation~\ref{equ:fluct_hydro_fluid}
in terms of $\Phi$ as 
\begin{eqnarray} 
\label{equ:Phi_fluct_hydro} \nonumber
\rho\Delta_{LB}\frac{\partial \Phi(t)}{\partial{t}} = (-\star\mb{d})
\mathcal{L}_f (-\star\mb{d}) -\star\mb{d}\mb{b}^{\flat} \\ \nonumber = \mu_m
\Delta_{LB}^2 \Phi -2\star\mb{d}\mu_m K(-\star\mb{d})\Phi \\ \nonumber -\star
\mb{d} \mathcal{T}_f (-\star \mb{d}) \Phi -\star\mb{d} \mb{b}^{\flat}
\hspace{1.4cm} \\ = \Delta_{LB} \left(\tilde{\mathcal{L}}_f \Phi + c\right).
\hspace{2.3cm} 
\end{eqnarray}
The $\mb{b}^{\flat} = \mbox{curl}_{\mathcal{M}}(c) = -\star \mb{d} c$ denotes
the contributions of the forces acting on the fluid.  
The $\Delta_{LB} = -\bs{\delta}\mb{d}$
denotes the Laplace-Beltrami operator~\cite{Abraham1988}.  The $\mathcal{T}_f$
denotes the traction stress operator of equation~\ref{equ_tract_stress_tplus}
and~\ref{equ_tract_stress_tminus}.  The $\mathcal{L}_f$ denotes the operator in
equation~\ref{equ:Lf_operator} for the hydrodynamic response.

We obtained equation~\ref{equ:Phi_fluct_hydro} by substituting for $\Phi$, using
$\mb{v}^{\flat} = -\star\mb{d}\Phi$, and using on both sides the generalized 
$\mbox{curl}_{\mathcal{M}}(\cdot) = -\star \mb{d}(\cdot)$.  We also 
use that the generalized curl commutes
with the operator $\mathcal{L}_f$ in the sense $(-\star \mb{d}) \mathcal{L}_f
(-\star \mb{d}) = \Delta_{LB} \tilde{\mathcal{L}}_f$.  The 
$\mathcal{\tilde{L}}_f$ takes on a form similar to
equation~\ref{equ:Lf_operator} but now applied to scalar fields.  We provide
more details for $\mathcal{\tilde{L}}_f$ in terms of spherical harmonic
coefficients below.  These considerations allow for the fluctuating
hydrodynamics to be expressed as \begin{eqnarray}
\label{equ:Phi_fluct_hydro_simplified} \nonumber \rho\frac{\partial
\Phi(t)}{\partial{t}} = \tilde{\mathcal{L}}_f \Phi + c.  \hspace{2.3cm}
\end{eqnarray}

In the overdamped regime, we need to compute the mobility tensor $M(\mb{X})$ of
the hydrodynamic coupling in equation~\ref{equ_Mobility_slip}.  This requires
us to solve the steady-state hydrodynamic equations~\ref{equ_steady_hydro}.
This can be reformulated as \begin{eqnarray} \label{equ:Phi_steady_hydro}
\nonumber \mu_m \Delta_{LB}^2 \Phi -2\star\mb{d}\mu_m K(-\star\mb{d})\Phi
\hspace{0.55cm} \\ -\star \mb{d} \mathcal{T}_f (-\star \mb{d}) \Phi =
\star\mb{d} \mb{b}^{\flat}.  \end{eqnarray}
We again use that the generalized
curl commutes with the operator $\mathcal{L}_f$ in the sense $(-\star \mb{d})
\mathcal{L}_f (-\star \mb{d}) = \Delta_{LB} \tilde{\mathcal{L}}_f$.  This
yields for the steady-state hydrodynamics \begin{eqnarray}
\label{equ:Phi_steady_hydro2} \nonumber \tilde{\mathcal{L}}_f \Phi = -c.
\hspace{2.3cm} \end{eqnarray}

Central to both the inertial fluctuating hydrodynamics and steady-state regimes
is computation of the action of the operator $\tilde{\mathcal{L}}_f \Phi$ which
gives the hydrodynamic responses of the system.  We expand $\Phi$ using
spherical harmonics to represent the vector potential as $\Phi = \sum_s \Phi_s
= \sum_s a_s Y_s$ where $\Phi_s = a_s Y_s$ and $s = (\ell,m)$ denotes the
spherical harmonics index, see Appendix~\ref{appendix:spherical_harmonics}.
For the spherical geometry, we can express solutions using that the spherical
harmonics are eigenfunctions of the Laplace-Beltrami
operator~\cite{AtzbergerSoftMatter2016,AtzbergerGrossHydro2018}.  

The action of the operator can be expressed as \begin{eqnarray}
\mathcal{\tilde{L}}_f \Phi_s = \Delta_{LB}^{-1}(-\star \mb{d})\mathcal{L}_f
(-\star\mb{d}) \Phi_s = L_s a_s Y_s, \end{eqnarray} where \begin{eqnarray}
\label{equ:L_s_hetero_mu} L_s & = & \frac{\mu_m}{R^2}\left[ 2 - \ell(\ell + 1)
\right. +  \hspace{3.15cm}   \\ && \nonumber \vspace{2cm} \left.  \left.
- \frac{R}{L^{+}}(\ell + 2)  
- \frac{R}{L^{-}}(\ell - 1) \right) \right], \end{eqnarray} with $L^{\pm} =
  \mu_m/\mu_{\pm}$.  
In the case with $\mu_{\pm} = \mu_f$ this can be
simplified to \begin{eqnarray} \label{equ:L_s_same_mu}   L_s & = &
\frac{\mu_m}{R^2} \left[ 2 - \ell(\ell + 1) 
- \frac{R}{L}\left(\ell + \frac{1}{2}\right) \right] \hspace{0.75cm}
  \end{eqnarray} where $L = \mu_m/(\mu_+ + \mu_-) = \mu_m/2\mu_f$.  We see from
this that the hydrodynamic flow on the sphere is characterized by the
non-dimensional parameter $\Pi_1 = L/R$, where $L = \mu_m/2\mu_f$ is the
Saffman-Delbr\"{u}ck length-scale~\cite{SaffmanDelbruck1975} and $R$ the radius
of the sphere. 

In the overdamped regime, to obtain the steady-state hydrodynamics, we can solve
the hydrodynamic equations to obtain directly the spherical harmonics
coefficients as 
\begin{eqnarray} 
\label{equ_Stokes_SPH_sol2_defA_ells} 
a_s & = & L_s^{-1} c_s.  
\end{eqnarray}

For the inertial fluctuating hydrodynamics of
equation~\ref{equ:fluct_hydro_fluid} and equation~\ref{equ:Phi_fluct_hydro}, we
can express the dynamics in terms of the spherical harmonics coefficients $a_s$
as \begin{eqnarray} \label{equ:Phi_fluct_hydro_sph_harmonics} \nonumber
\atzEquNumAndAdd \\ \nonumber \frac{\partial a_s(t)}{\partial{t}} = \rho^{-1}
L_s a_s(t) + \rho^{-1}\bar{c}_s(t) + g_s(t) + h_s(t), \end{eqnarray} where $c_s
= \bar{c}_s + g_s + h_s$.  The stochastic driving terms $\{g_s\}$ are
associated with the hydrodynamics.  These are complex-valued Gaussian processes
$\delta$-correlated in time with mean zero and covariance 
\begin{eqnarray}
\label{equ:Phi_fluct_hydro_sph_harmonics_stoch_g} \nonumber \atzEquNumAndAdd \\
\nonumber \left\langle g_s(t)\overline{g_{s'}(r)} \right\rangle = -2k_B{T}
\rho^{-2} |\lambda_s|^{-1} L_s \delta_{ss'} \delta(t - r).  
\end{eqnarray}
The
stochastic driving terms $\{h_s\}$ are associated with the fluid-structure
coupling discussed in Section~\ref{sec:IB_curved}.  This involves 
the contributions from the dissipative terms 
in equation~\ref{equ:fluct_hydro_fluid}.  
The
$h_s$ are complex-valued Gaussian processes having correlations with the
microstructure stochastic dynamics in
equation~\ref{equ:fluct_hydro_microstructure}. We can express this as
\begin{eqnarray} \label{equ:Phi_fluct_hydro_sph_harmonics_stoch_h} h_s(t) =
\hspace{6.1cm} \\ \nonumber \hspace{0.1cm} -\rho^{-1}\left\langle
\Delta_{LB}^{-1}(-\star\mb{d}) \Lambda[\mb{F}_{thm}(t)](\mb{x}), Y_s(\mb{x})
\right\rangle_{L^2}.  \end{eqnarray} 
The $\langle \cdot,\cdot \rangle_{L^2}$ gives the surface $L^2$-inner product.
The $\lambda_s = -\ell(\ell + 1)/R^2$
denotes the $s^{th}$ eigenvalue of the Laplace-Beltrami operator $\Delta_{LB} =
-\bs{\delta}\mb{d}$. The $\delta_{a,b}$ denotes the Kronecker $\delta$-function
which is zero for $a \neq b$ and one for $a = b$.   

The real-space fluctuating velocity field $\mb{v}^{\flat}$ can be recovered
using $\mb{v} = (-\star \mb{d} \Phi)^{\sharp}$.  This is derived from the
stochastic force $\mb{F}_{thm}$ acting on microstructures in
equation~\ref{equ:fluct_hydro_microstructure}.  Together, these stochastic
driving terms give in real-space $\mb{f}_{thm}^{\flat} = -\star \mb{d} \sum_s
\rho (g_s + h_s) Y_s = \mb{g}_{thm}^{\flat} - \Lambda[\mb{F}_{thm}]$ which
satisfy equations~\ref{equ_stoch_driving_fluid}--\ref{equ_stoch_driving_cross}.

In summary, the
equations~\ref{equ:Phi_fluct_hydro_sph_harmonics}--
~\ref{equ:Phi_fluct_hydro_sph_harmonics_stoch_h}
give a formulation for surface fluctuating hydrodynamics incorporating
fluid-structure interactions in terms of the vector potential $\Phi$.  The use
of such Hodge decompositions and vector potentials is particularly convenient
when representing velocity fields on surfaces.  We give additional details on
the derivation of these results in
Appendix~\ref{appendix:derivation_fluct_hydro_phi}.

\section{Stochastic Numerical Methods} 
\label{sec:stoch_methods} We develop
numerical methods for the drift-diffusion dynamics of microstructures both in
the inertial and overdamped regimes.  To handle the incompressibility
constraint for the surface hydrodynamics we make use of the generalized vector
potential formulation discussed in
Section~\ref{sec:hydrodynamic_vec_poten_formulation}.

\subsection{Inertial Regime} 
In the inertial regime, we formulate the time-step
integration of the fluctuating hydrodynamics of
equation~\ref{equ:fluct_hydro_fluid} in terms of stochastic dynamics of the
vector potential $\Phi$ in equation~\ref{equ:Phi_fluct_hydro}
and~\ref{equ:Phi_fluct_hydro_sph_harmonics}.  We use
a time-step update in terms of the spherical harmonics 
coefficients of $\Phi$ given by
\begin{eqnarray}
\label{eqn:hydro_update}
\nonumber \atzEquNumAndAdd \\ \nonumber a_s^{n+1} = a_s^n +
\Delta{t}\rho^{-1}L_s a_s^n + \Delta{t}\rho^{-1}c_s^n + g_s^n + h_s^n.
\end{eqnarray}
For the microstructure dynamics, we use an update 
\begin{eqnarray} 
\label{eqn:verlet_update} \\ \nonumber \mb{V}^{n+1/2} &=&
\mb{V}^{n} - \frac{\Delta{t}}{2m}\gamma(\mb{V}^n - \Gamma^n \mb{v}^n) +
\frac{\Delta{t}}{2m} \mb{F}_{thm}^n\\ \nonumber & + &
\frac{\Delta{t}}{2m}\mb{F}^{n}\\ \nonumber \mb{X}^{n+1} &=& \mb{X}^{n} +
\Delta{t} \mb{V}^{n+1/2} \\ \nonumber \mb{V}^{n+1} &=& \mb{V}^{n} -
\frac{\Delta{t}}{2m}\gamma(\mb{V}^n - \nonumber \Gamma^n \mb{v}^n) +
\frac{\Delta{t}}{2m} \mb{F}_{thm}^n\\ \nonumber & + &
\frac{\Delta{t}}{2m}\mb{F}^{n+1}.  
\end{eqnarray} 
We have $\Phi^n = \sum_s
a_s^n Y_s$ with the fluid velocity obtained by $\mb{v}^n = (-\star \mb{d}
\Phi^n)^{\sharp}$.  Similarly, from the fluid-structure coupling $\mb{c}^n =
\Lambda^n[\gamma(\mb{V}^n - \Gamma^n \mb{v}^n]$ we obtain the terms $c_s^n =
\left\langle \Delta_{LB}^{-1}(-\star\mb{d}) \mb{c}^n,
Y_s(\mb{x})\right\rangle_{L^2}$.
The hydrodynamic time-step updates in 
equation~\ref{eqn:hydro_update} are 
related to the Euler-Maruyama method~\cite{Platen1992} 
and the microstructure updates
in equation~\ref{eqn:verlet_update} 
are related to modified velocity-verlet 
methods~\cite{Verlet1967,Frenkel2002_chMD,
Farago_Langevin_Num_Method_2013}.
This gives in
the stochastic setting errors with worse-case scalings
$O(\sqrt{\Delta{t}})$,~\cite{Verlet1967,Frenkel2002_chMD,
Farago_Langevin_Num_Method_2013}.

We account for thermal fluctuations of the microstructures over the time-step
through the Gaussian term $\mb{F}_{thm}^n$.  This has mean zero and covariance
\begin{eqnarray} \langle \mb{F}_{thm}^n \mb{F}_{thm}^m \rangle = 2k_B{T}\gamma
\Delta{t}^{-1} \delta_{m,n}.  \end{eqnarray} The $g_s^n$ are complex-valued
Gaussians with mean zero and covariance \begin{eqnarray} \left\langle g_s^{n}
g_{s'}^{m} \right\rangle = -2 \rho^{-1} L_s \mathcal{C}_{ss'} \Delta{t}
\delta_{nm}.  \end{eqnarray} The $L_s$ are given by
equation~\ref{equ:L_s_same_mu} or~\ref{equ:L_s_hetero_mu} and
$\mathcal{C}_{ss'}$ are the equilibrium fluctuations of the modes $a_s$ given
by \begin{eqnarray} \mathcal{C}_{ss'} = \left\langle a_s a_{s'} \right\rangle =
\rho^{-1} k_B{T} |\lambda_s|^{-1} \delta_{s',\overline{s}}.  \end{eqnarray} The
$\lambda_{s} = -{\ell(\ell + 1)}/{R^2}$ are the eigenvalues of the
Laplace-Beltrami operator $\Delta_{LB} = -\bs{\delta}\mb{d}$ on the surface.
For $s = (\ell,m)$ the $\overline{s} = (\ell,-m)$.

The $h_s^n$ are complex-valued Gaussians which we generate from
$\mb{F}_{thm}^n$ by \begin{eqnarray} \label{equ:h_s_n_numerical} h_s^n =
\hspace{6.75cm} \\ \nonumber \hspace{0.25cm} -\rho^{-1}\Delta{t}\left\langle
\Delta_{LB}^{-1}(-\star\mb{d}) \Lambda^n[\mb{F}_{thm}^n](\mb{x}), Y_s(\mb{x})
\right\rangle_{L^2}.  \end{eqnarray} This ensures proper correlations in the
system between the microstructure and the fluctuating hydrodynamics. We note
that in our derivation of $\mb{F}_{thm}$ there were negative cross-correlations
with the fluid which we capture in our numerical methods consistent with
equation~\ref{equ_stoch_driving_cross}.  These cross-correlations play the
important role of ensuring momentum conservation.  They account for the
spontaneous fluctuations that exchange momentum back and forth between the
fluid and microstructures.

The integrator approach we have introduced works well when the time-scales are
comparable between the microstructure evolution and hydrodynamics.   When there
is a disparity in these time-scales, stiff stochastic numerical time-step
integrators can also be developed using the vector potential formulation.  For
instance, we can develop exponential time-stepping approaches similar to our
prior works~\cite{AtzbergerSIB2007,AtzbergerOsmoticVesicles2015}.  As an
alternative, an asymptotic analysis of the stochastic dynamics of the
fluid-structure system can also be performed as in~\cite{AtzbergerTabak2015}.
This can be used to formulate equations in overdamped regimes and develop
stochastic numerical methods~\cite{AtzbergerTabak2015}.

\subsection{Overdamped Regime} 
For the overdamped regime, we develop numerical
methods for simulations based on stochastic variants of the velocity-verlet
method~\cite{Verlet1967,Frenkel2002_chMD,Farago_Langevin_Num_Method_2013}.
We account for the thermal drift
term in the stochastic dynamics using an approach related to methods 
in~\cite{DonevStochIntegrator2014,Fixman1978}. We update the collective
configuration $\mb{X}$ of the particles or microstructures of the system using
\begin{equation} \label{equ_stoch_integrator} \begin{split} \mb{V}^n &=
M(\mb{X}^n)\mb{F}^n + Q(\mb{X}^n)\boldsymbol{\xi}^n  \\ \mb{\tilde{X}}^{n + 1}
&= \mb{X}^{n} + \mb{V}^n \Delta{t} \\ \tilde{\mb{V}}^{n+1} &=
M(\mb{\tilde{X}}^{n+1})\mb{\tilde{F}}^{n+1} + Q(\mb{X}^n)\boldsymbol{\xi}^n \\
\mb{X}^{n + 1} &=  \frac{1}{2}\left(\mb{V}^{n} +
\tilde{\mb{V}}^{n+1}\right)\Delta{t} \\ + k_BT&
\left(\frac{\Delta{t}}{\delta}\right)\left\langle \left(M\left(\mb{X}^n +
\delta \boldsymbol{\hat{\xi}}\right)
- M\left(\mb{X}^n\right)\right)\boldsymbol{\hat{\xi}} \right\rangle_{\bar{N}}.
  \end{split} \end{equation}

The thermal fluctuations have correlations generated 
by ${Q}(\mb{X}^{n})$, where
${Q}(\mb{X}^{n}){Q}(\mb{X}^{n})^{T} = 2 k_B{T}M(\mb{X}^{n})/\Delta{t}$ 
$\boldsymbol{\xi}$. The $\bs{\xi}$ are standard Gaussian random 
variates with independent
components having mean zero and variance one.  The thermal drift $k_B{T}
\nabla\cdot \mb{M}$ is approximated by numerically estimating an average by
$\langle Z \rangle_{\bar{N}} = 1/\bar{N} \sum_{k=1}^{\bar{N}} Z^{[k]}$.  For
the random variable $Z$, the $Z^{[k]}$ denotes one of the $\bar{N}$ independent
samples.  
This provides a probabilistic estimator for the divergence.
The random variables $\mb{p}$ and $\mb{q}$ satisfy $\langle {p}_i
{q}_j \rangle = \delta_{ij}$.  This gives
\begin{eqnarray} 
\\ 
\nonumber
\lim_{\delta\rightarrow 0} {\delta}^{-1} \left\langle \left(M\left(\mb{X} +
\delta\mb{p}\right) - M(\mb{X})\right)\mb{q} \right\rangle = \nabla\cdot M.
\end{eqnarray} 
This follows an approach related to~\cite{DonevStochIntegrator2014,Fixman1978}.
The validity follows readily by Taylor
expanding $M$ in the variable $\mb{X}$~\cite{Goldberg1976}.

The stochastic velocity-verlet scheme in equation~\ref{equ_stoch_integrator}
can be viewed in stages as follows.  The $\mb{\tilde{X}}^{n+1}$ gives the
predictor part of the update of the configuration that is used to evaluate the
force $\mb{\tilde{F}}^{n+1}$ and mobility $M(\mb{\tilde{X}}^{n+1})$.  The
$\mb{X}^{n+1}$ gives the corrector part for the update of the configuration
which makes use of the predictor data and additional contributions from the
thermal drift term.

\subsubsection{Generating Stochastic Forces from the Mobility Tensor} During
numerical time-step integration of the drift-diffusion dynamics by
equation~\ref{equ_stoch_integrator}, we must generate the stochastic driving
term $\mb{h}^n = Q(\mb{X}^n) \bs{\xi}^n$.  One way to do this is to
determine from the mobility the square-root factor $Q$ satisfying
$Q(\mb{X}^n)Q(\mb{X}^n)^T = 2k_B{T}M$.  Cholesky factorization methods
can be used for symmetric positive definite
matrices~\cite{Trefethen1997}. However, in the current formulation this present
challenges given that the mobility tensor for the surface hydrodynamics is
singular since it only depends on the tangential component of the force, see
equation~\ref{equ:Phi_fluct_hydro}.  Working only with the surface coordinates
also presents challenges since there is no global coordinate chart for 
spherical topologies~\cite{AtzbergerSoftMatter2016}.
We instead work in the embedding space, but this results in a 
representation for the mobility tensor that is only positive 
semi-definite preventing direct
use of the Cholesky factorization method. 

We remedy this situation by adding a block diagonal 
tensor to our grand mobility tensor $M$ of 
equation~\ref{equ_Mobility_no_slip}.  We do this in a way
that preserves the drift-diffusion dynamics in the 
tangential directions.  This is accomplished by using
rank-one stabilizations to obtain modified self-mobility  
blocks of the form $\tilde{M}_{ii} =
M_{ii} + \alpha\mb{n}_i\mb{n}_i^T$.  The $\alpha>0$ is 
any chosen positive weight.  
The $\mb{n}_i = \mb{n}(\mb{X}_i)$ 
denotes the normal vector on the sphere at the location of
$\mb{X}_i$.  By using this modification, we can obtain a tensor that
is positive definite.  This allows us to use Cholesky
factorizations to obtain $\tilde{Q}$ satisfying
$\tilde{Q}(\mb{X}^n)\tilde{Q}(\mb{X}^n)^T = 2k_B{T}\tilde{M}$. 

Our introduced stabilizations are constructed orthogonal to the
tangent directions of the surface.  As a result, this only modifies 
the mobility responses in the directions normal to the surface and 
preserves the microstructure drift-diffusion dynamics in the tangential 
directions.  We generate the tangential stochastic forces using
$\mb{h} = \wp \tilde{\mb{h}}^n$, where $\tilde{\mb{h}}^n = Q(\mb{X}^n)
\bs{\xi}^n$.  The $\bs{\xi}^n$ are Gaussian vectors with independent components
having mean zero and variance one.  The $\wp$ denotes projection of the
stochastic vector $\tilde{\mb{h}}^n$ to the tangent space of the
surface.  This allows us to generate the stochastic terms for
the thermal fluctuations in the surface drift-diffusion dynamics of
the microstructures.

\subsection{Computing Hydrodynamic Forces and Responses}
\label{sec:hydrodynamic_forcing}

To solve for the fluid velocity, we represent the force density term as $c =
\mbox{curl}_\mathcal{M}(\mb{b}^{\flat}) = 
-\star \mb{d} \mb{b}^{\flat}$ with expansion $c = \sum_\ell c_s Y_s$.  We
compute this in practice by using the spherical harmonics representation of
$\mb{b}$ obtained by $\hat{b}^{[a]}_s = \langle {b}^{[a]}, Y_s \rangle$.
The $\mb{b} = {b}^{[a]}
\partial_{z^a}$ gives the representation of the force density $\mb{b}$ in the
ambient physical space with coordinates $z^a$. 
The inner-products $\langle \cdot,\cdot \rangle$ are 
approximated numerically by using Lebedev
quadrature~\cite{Lebedev1976,Lebedev1999}. 
We then compute 
$c = \mbox{curl}_\mathcal{M}(\mb{b}^{\flat}) = -\star
\mb{d} \mb{b}^{\flat} = \sum_s (-\star \mb{d}) \mb{b}_s$.  This uses the finite
spherical harmonics expansion of $\mb{b} = \sum_s \mb{b}_s$, where $\mb{b}_s$
denotes the term of the expansion associated with index $s$.  We compute $c_s$
using ${c}_s = \langle c, Y_s \rangle$, where the inner-product is approximated
numerically using Lebedev quadrature~\cite{Lebedev1976,Lebedev1999}.  From the
solution coefficients $a_s$, we obtain the fluid velocity on the surface from
$\mb{v} = \sum_s a_s (-\star \mb{d}Y_s)^{\sharp}$.

In practice, we numerically approximate $L^2$ inner-products by using Lebedev
quadratures on the spherical
surface~\cite{Lebedev1976,Lebedev1999,AtzbergerSoftMatter2016}. We use $\langle
u,v \rangle = \langle u, v \rangle_Q = \sum w_i u(\mb{x}_i)v(\mb{x}_i)$, where
$\langle u, v \rangle_Q$ denotes the quadrature computed inner-product with
$w_i$ the weights and $\mb{x}_i$ the
nodes~\cite{Lebedev1976,Lebedev1999,AtzbergerSoftMatter2016}.  
This provides a
finite spherical harmonics expansion of 
$\mbox{curl}_\mathcal{M}(\mb{b}) = -\star \mb{d} \mb{b}$.  
While other
quadrature approaches could be used such as latitude and longitude sampling,
the Lebedev quadrature nodes have octahedral symmetry and provide a better
distribution of sampling nodes on the spherical
surface~\cite{AtzbergerGrossHydro2018,AtzbergerSoftMatter2016}.

We remark that for the hydrodynamics on the sphere most of the 
representation of the flow responses and evolution is represented 
analytically using the spherical harmonics expansions.  As a
result, in the developed numerical methods the primary 
source of errors is from the truncation 
of the spherical harmonics expansion and related projections.
In these methods, 
the primary source of computational expense arises 
from computing the $L^2$-inner products
using the quadratures and in summing the spherical
harmonics expansions in reconstructions.

When using $n$ modes, this part of the 
numerical algorithms have a computational 
complexity of $O(n^2)$.  In the inertial 
regime, the thermal fluctuations
requires initially a single 
off-line computation of the Cholesky 
factorization of the covariance which costs $O(n^3)$.
Generating the correlated variates then 
costs $O(n^2)$ each time-step.
In the 
overdamped case, the primary additional 
computational expense is from the Cholesky 
factorization of the mobility tensor.  
For $m$ microstructure
degrees of freedom and $n$ modes, 
the overdamped case has computational complexity 
$O(n^2 + m^3)$.

The Lebedev quadratures 
allow for computing the $L^2$-inner products to a high 
level of accuracy.  Available Lebedev quadratures 
allow for $L^2$ inner-products for functions
expanded up to degree $65$ spherical
harmonics to be computed up to round-off errors
~\cite{Lebedev1999}.
We performed convergence analysis of our 
spectral numerical methods using Lebedev quadratures 
for approximating 
exterior calculus operators and solving hydrodynamic equations
in~\cite{AtzbergerGross2017,AtzbergerGrossHydro2018}.

\section{Applications} \label{sec:applications}

We discuss a few applications showing how 
to use our introduced fluctuating hydrodynamics
methods for investigating phenomena within curved fluid
interfaces.  We first discuss the hydrodynamic 
relaxation of fluctuating fluids
and characterize scalings of the velocity autocorrelations.  We find
different scalings can emerge depending on the physical regimes associated with
the interface geometry, surface viscosity, and bulk viscosity.  We further
investigate the mobilities associated with microstructures embedded within
fluid interfaces in the quasi-steady regime.  We then demonstrate the
computational methods by studying the correlated diffusion of passive particles
and the drift-diffusion dynamics of active microswimmers.  The results show
some of the rich phenomena within curved fluid interfaces that 
can arise from hydrodynamic coupling, thermal 
fluctuations, and geometry that our methods can be used to investigate.

\subsection{Autocorrelations of the Surface Fluctuating Hydrodynamics}
\label{sec:autocorrelation_fluct_hydro}

We investigate the autocorrelation of the fluid velocity of interfacial
fluctuating hydrodynamics introduced in equation~\ref{equ:fluct_hydro_fluid}.
We find the velocity autocorrelations can exhibit power-law decay with scalings
$\tau^{-1}$ and $\tau^{-2}$ depending on the physical regime.  We also find in
some regimes a plateau behavior can arise.  This differs from
bulk three dimensional fluctuating hydrodynamics that would exhibit 
$\tau^{-3/2}$ power-law decay.  This also differs from purely 
two dimensional fluctuating hydrodynamics that would exhibit
only a $\tau^{-1}$ power-law decay~\cite{AtzbergerVelCorr2006,AtzbergerSIB2007}.

Interfacial fluctuating hydrodynamics involves both dissipation from the
propagation of shearing motions within the interfacial fluid surface and from
traction coupling with the bulk surrounding fluid.  This results in two
important time-scales.  The first is the time-scale for shear stresses to
propagate over the entire spherical surface $\tau_f = R^2/\mu_m$.  The second
is the time-scale on which rigid-body rotation of the entire spherical
interface dissipates energy significantly to the bulk surrounding fluid $\tau_r
= RL/\mu_m$.  

\end{multicols} \begin{figure}[H] \centering
\includegraphics[width=0.85\columnwidth]
{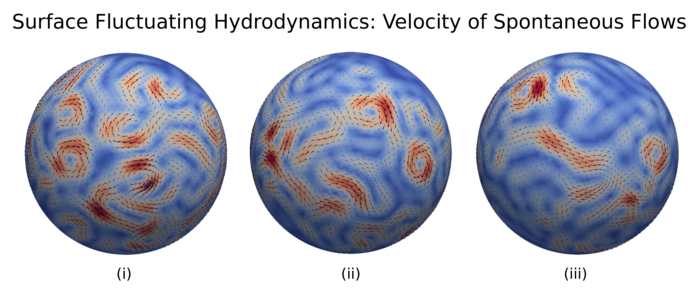}
\caption{Surface Fluctuating Hydrodynamics and Velocity of Spontaneous Flows.
We show a few samples of the fluid velocity of the surface fluctuating
hydrodynamics.  We emphasize these results have an introduced correlation
length-scale by use of a finite spherical harmonics expansion with modes up to
degree $N = 20$.  In the drift-diffusion dynamics of microstructures, similar
correlations arise from the averaging operator $\Gamma$.  The surface
fluctuating hydrodynamics incorporates the dissipation from both the
interfacial shear viscosity and the traction stresses with the inner and outer
bulk surrounding fluids using equation~\ref{equ:Lf_operator}.  The
incompressibility of the fluid and the spherical topology induces long-range
correlated structures that manifest as recirculation patterns on the surface.
} \label{fig:SurfaceFluctHydro} \end{figure} \begin{multicols}{2}

\begin{figure}[H] \centering
\includegraphics[width=1.0\columnwidth]
{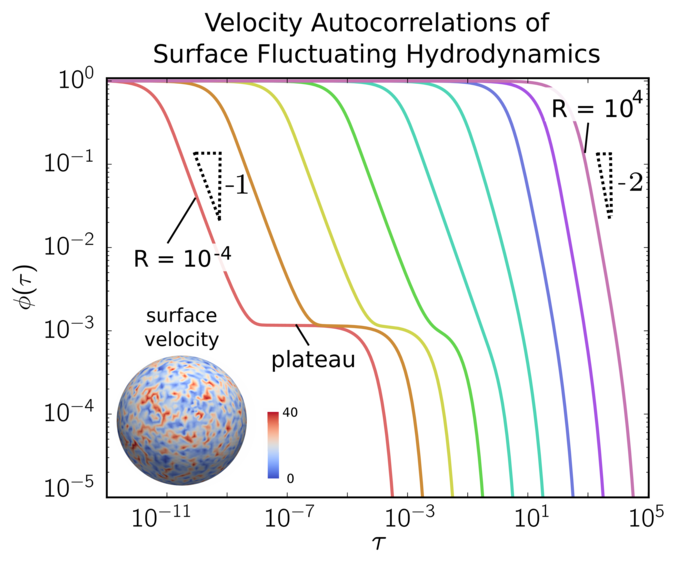}
\caption{Velocity Autocorrelations of Surface Fluctuating Hydrodynamics.  For
the interfacial fluctuating hydrodynamics introduced in
equation~\ref{equ:fluct_hydro_fluid} we show the temporal autocorrelation
$\phi(\tau) = \langle \mb{v}(0)\cdot \mb{v}(t)\rangle/\langle \mb{v}^2(0)
\rangle$. The curves left to right have $R = 10^{\alpha}$ with integer values
in range $\alpha \in [-4,4]$.  We see the velocity autocorrelation  exhibits
regimes with power-law decay $\tau^{\beta}$ with $\beta = -1,-2$ (dotted lines)
and plateaus.  The surface fluctuating hydrodynamics where sampled with
truncated expansion with modes up to degree $N = 50$.  We derive these
power-laws for surface fluctuating hydrodynamics in
Appendix~\ref{appendix:autocor_v}.   } \label{fig:FluctHydroAutoCor}
\end{figure}

We explore these contributions to the velocity autocorrelations by varying the
ratio $\tau_f/\tau_r = R/L$.  This involves the sphere radius $R$ and the
Saffman-Delbr\"{u}ck length $L$.  We show our results in
Figure~\ref{fig:FluctHydroAutoCor}.  We take as default parameter values $\rho
= 1$, $\mu_m = 1$, $K_B{T} = 1$, $R = 1$, $L = 1$, and $\ell \leq N=\ell_* =
50$.   We give details on our derivations analyzing these autocorrelations in
Appendix~\ref{appendix:autocor_v}. 

The velocity autocorrelations of the surface fluctuating hydrodynamics can be
expressed using the following expansion in spherical harmonics
\begin{eqnarray}
\label{equ:expand_autocor_main} \left\langle v^{\phi}(0) v^{\phi}(t)
\right\rangle \hspace{5cm} \\ \nonumber \hspace{1cm} = \sum_s \exp(tL_s)
\frac{k_B{T}}{|g|\rho \ell(\ell + 1)} \left|\frac{\partial Y_s}{\partial
\theta}\right|^2 \\ \nonumber \hspace{1cm}\approx C \frac{k_B{T}}{|g|\rho}
\sum_\ell  \left(\ell + \frac{1}{2}\right) \exp(tL_{\ell}).  
\end{eqnarray} 
The $v^{\phi}$ denotes the $\phi$-directional component of the velocity and the $Y_s$
denotes the spherical harmonic with mode $s = (\ell,m)$.  The metric factor on
the sphere is $|g| = R^2$.  We also use that the velocity is isotropic and that
$\langle a_s^2 \rangle = k_{B}T R^2/\rho \ell(\ell + 1)$ and $\langle a_s(0)
a_{s}(t) \rangle = \langle a_s^2 \rangle \exp(tL_s)$, where $L_s$ is given in
equation~\ref{equ:L_s_same_mu}.   We derive scaling laws for different physical
regimes using equation~\ref{equ:expand_autocor_main}.  We remark that all of
our results use a truncated expansion with modes up to degree $N=50$.   This
introduces a regularization length-scale similar to our immersed boundary
coupling approaches for determining responses of particles and microstructures.
Details of our derivations can be found in Appendix~\ref{appendix:autocor_v}.     

When $R/L \ll 1$, we find the velocity correlations exhibit an initial decay to
a power-law regime with scaling $\tau^{-1}$ on time-scales $\tau \ll \tau_f$.
This is followed by a plateau regime that persists from $\tau_f \ll \tau \ll
\tau_r$.  This eventually gives way to exponential decay for $\tau \gg \tau_r$.
The initial $\tau^{-1}$ power law decay is associated with the propagation of
shear stresses over the interface with negligible dissipation to the bulk.
This persists until time-scale $\tau_f$,  see
Figure~\ref{fig:FluctHydroAutoCor}.  

We remark that this is similar
to the velocity autocorrelations that would be observed in a purely 
two dimensional viscous Newtonian fluid, which can be computed 
readily using the methods in~\cite{AtzbergerVelCorr2006}.  In 
contrast, for the spherical fluid interface the two dimensional 
fluid has finite area.  This results in finite size effects 
in the flow responses and correlations.

A plateau arises in the case when $\tau_r \gg \tau_f$ which creates an
intermediate regime.  In the intermediate plateau regime, the shear
stress has already propagated to the entire surface, but the dissipation into
the bulk fluid of the rigid-body rotational motion has not yet become
significant.  On time-scales $\tau \gg \tau_r$, the dissipation into the bulk fluid
dominates through the rotational motions and gives exponential decay.  As the
$R/L$ increases, the plateau regime disappears when $\tau_r \sim \tau_f$, as
seen when moving left to right in Figure~\ref{fig:FluctHydroAutoCor}.

In the $R/L \gg 1$ regime, the time-scales for decay from intra-interface
shearing motions becomes reversed and large relative to the time-scale for
decay from the coupling to the bulk fluid $\tau_f \gg \tau_r$.  This results in
a new regime with power law scaling $\tau^{-2}$ for time-scales $\tau \ll
\tau_r$.  When $\tau \gg \tau_f$ this again eventually gives exponential decay 
from the rotational motions of the entire interface.  The $\tau^{-2}$ arises from
simultaneous dissipation from the shearing motions within the interface and
dissipation from coupling to the bulk fluid.  We give more details and
derivations of these power-laws and related time-scales in
Appendix~\ref{appendix:autocor_v}.

\subsection{Mobility Tensor for Interacting Particles and Microstructures}
\label{sec:mobility} 
We compute numerically the mobility 
$\tilde{M}(\mb{X}) = \Gamma
\tilde{\mathcal{S}} \Lambda$ using 
equation~\ref{equ_Mobility_no_slip}.  We obtain a numerical solver
$\tilde{\mathcal{S}}$ for the fluid velocity $\tilde{\mb{v}}^{\flat} =
\tilde{\mathcal{S}} \Lambda[\mb{F}]$ with force distribution
$\mb{f}^{\flat}(\mb{x}) = \Lambda[\mb{F}](\mb{x})$ on the spherical surface
using equation~\ref{equ:Phi_steady_hydro} and Lebedev
quadratures~\cite{Lebedev1976,Lebedev1999,AtzbergerSoftMatter2016}.
We performed convergence analysis of the numerical methods 
based on Lebedev spectral approaches 
in~\cite{AtzbergerGross2017,AtzbergerGrossHydro2018}.

From the symmetry of the sphere the mobility of $N$ particles can be determined
from the canonical two particle mobility tensor.  The two particle mobility
obtained from equation~\ref{equ_Mobility_no_slip} can be expressed as
\begin{eqnarray} \label{equ:mobility_two} M = \left[ \begin{array}{ll} M_{11} &
M_{12} \\ M_{21} & M_{22} \\ \end{array} \right] \end{eqnarray} with
\begin{eqnarray} \label{equ:mobility_two} \bar{M}_{12} = \left[
\begin{array}{ll} M_{\parallel\parallel} & M_{\parallel\perp} \\
M_{\perp\parallel} & M_{\perp\perp} \\ \end{array} \right].  \end{eqnarray} The
$M_{ij} = \Gamma_i \tilde{\mathcal{S}} \Lambda_j$ where we use notation
$\Gamma_i\mb{v}^{\flat} = \Gamma[\mb{X}_i] \mb{v}^{\flat}$ and
$\Lambda_j[\mb{F}](\mb{x}) = \Lambda[\mb{X}_j][\mb{F}](\mb{x})$.  The $M_{ii}$
denote the self-mobilities and can be computed numerically once and stored.

\begin{figure}[H]
\includegraphics[width=1.0\columnwidth]{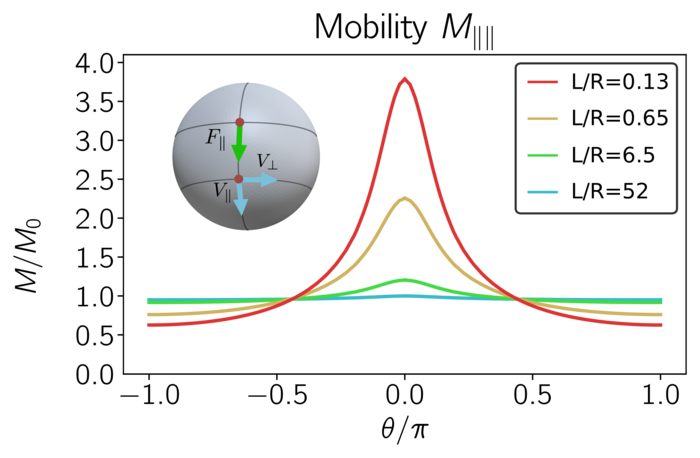}
\caption{Mobility Response $\mb{V}_{\parallel} = M_{\parallel\parallel}
F_{\parallel}$.  We show the mobility response component when a unit force is
applied to the first particle in the direction parallel to the separation of
the two particles.  We scale the mobility by $M_0 = 641$ corresponding to a
pure rotational response.  The response $\mb{V}_{\perp} = M_{\perp\parallel}
F_{\parallel}$ was found to be negligible with a magnitude smaller than $0.01\%$
of $M_0$.  } \label{fig:Mobility_M_para} \end{figure}

\begin{figure}[H]
\includegraphics[width=1.0\columnwidth]{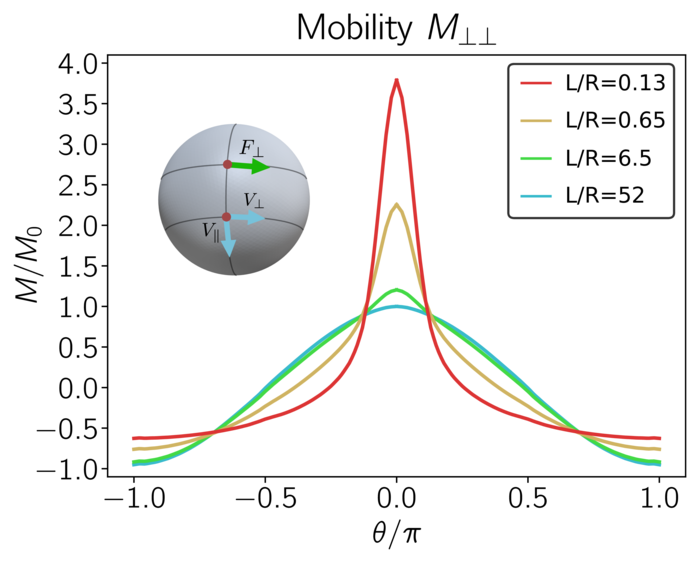}
\caption{Mobility Response $\mb{V}_{\perp} = M_{\perp\perp} F_{\perp}$.  We
show the mobility response component when a unit force is applied to the first
particle in the direction perpendicular to the separation of the two particles.
We scale the mobility by $M_0 = 641$ corresponding to a pure rotational
response.  The response $\mb{V}_{\parallel} = M_{\parallel\perp} F_{\perp}$ was
found to be negligible with a magnitude smaller than $0.01\%$ of $M_0$.  }
\label{fig:Mobility_M_perp} \end{figure}

The $\bar{M}_{12}$ give for a force $\mb{F}$ on the first particle the velocity
response $\mb{V}$ at the second particle.  We can express this as $\mb{V} =
V_{\parallel}\mb{e}_{\parallel} + V_{\perp}\mb{e}_{\perp}$, where for the two
particle displacement we split into the parallel $\parallel$ and perpendicular
$\perp$ components.  The $V_{\parallel} =
M_{\parallel\parallel}\mb{F}_{\parallel} + M_{\parallel\perp}\mb{F}_{\perp}$
and $V_{\perp} = M_{\perp\parallel}\mb{F}_{\parallel} +
M_{\perp\perp}\mb{F}_{\perp}$.  Using the symmetry of the sphere we can
tabulate numerically the two particle mobility tensor by using a canonical
configuration of the two particles.  We rotate the sphere so that the first
particle is always situated at the north pole.  We then align the geodesic
displacement between the two particles in the $xz$-plane with tangent along the
positive $x$-axis.  We denote the rotation operation by $\mathcal{R}^T$ that
moves any two particles into this canonical configuration $\tilde{\mb{X}} =
\mathcal{R}^T \mb{X}$.  We can convert our canonical tabulated two-particle
mobilities to the specific mobility of two particles by $M_{12} = \mathcal{R}
\tilde{M}_{12} \mathcal{R}^T$.  We obtain the grand-mobility tensor $M$ for $n$
interacting particles by summing over all pairs of particles the two particle
mobility tensors.  We show the components of the two-particle mobility
$\bar{M}_{12}$ when $L/R = 0.13, 0.65, 6.5, 52$ in
Figure~\ref{fig:Mobility_M_para} and~\ref{fig:Mobility_M_perp}.

\subsection{Equilibrium Fluctuations} \label{sec:stat_mech} We validate our
stochastic numerical methods by studying the drift-diffusion dynamics of
interacting particles.  A strong indication of the validity the methods is
provided by consider how particles diffuse over time when subject to a
conservative force.  This requires the stochastic methods to capture both the
drift dynamics accurately while also handling appropriately the thermal
fluctuations of the system.  From equilibrium statistical mechanics a
configuration should have a probability distribution $\rho(\mb{X})$ of the
Gibbs-Boltzmann form~\cite{Reichl1998,Chandler1987}   \begin{eqnarray}
\label{equ_equil_distr_analytic} \rho(\mb{X}) = \frac{1}{Z}
\exp\left[-\phi(\mb{X})/{k_B{T}}\right].  \end{eqnarray} The $Z$ denotes the
partition function and $k_B{T}$ denotes the thermal energy of the
system~\cite{Reichl1998}.

\begin{figure}[H]
\centerline{\includegraphics[width=1.0\columnwidth]
{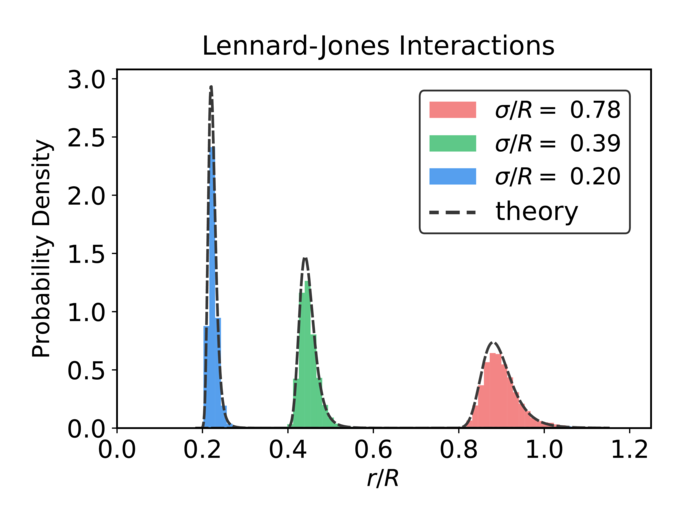}}
\caption{Gibbs-Boltzmann Distribution.  We compute using our stochastic
numerical methods of equation~\ref{equ_stoch_integrator} the equilibrium
fluctuations associated with the drift-diffusion motions of hydrodynamically
coupled particles having Lennard-Jones interactions of equation~\ref{equ_lj}.
We test interactions for a few choices of $\sigma$ and find good agreement
between our numerical results and the Gibbs-Boltzmann distribution predicted by
equation~\ref{equ_equil_distr_analytic}. } \label{fig:validation_lj}
\end{figure}

We consider the drift-diffusion dynamics of two hydrodynamically coupled
particles having the non-linear Lennard-Jones interaction \begin{eqnarray}
\label{equ_lj} \phi(r) = 4\epsilon\left(\left(\dfrac{\sigma}{r}\right)^{12} -
\left(\dfrac{\sigma}{r}\right)^6\right).  \end{eqnarray} The $r$ denotes the
distance between the two particles.  The $\sigma$ denotes the length-scale
characterizing the radius of the particles.  We find our stochastic
numerical methods provide for the drift-diffusion dynamics very good agreement
with the distribution predicted from equilibrium statistics mechanics from
equation~\ref{equ_equil_distr_analytic}.  We show these results for a few
choices of $\sigma$ in Figure~\ref{fig:validation_lj}.    

For parameters, we take throughout the thermal energy $k_BT =
2.48$\hspace{0.075cm}amu$\cdot$nm$^2$/ps$^2$, the strength of the potential
$\epsilon = 10K_BT$, viscosity ratio $L/R = 0.65$.  We use for the stochastic
integrator in equation~\ref{equ_stoch_integrator} the time-steps $\Delta{t} =
1.3\times 10^{5}$ ps and drift estimator $\delta = 10^{-1}$, $\bar{N} = 10$.

\subsection{Hydrodynamic Correlations in Particle Diffusion}

Motivated by proteins in lipid bilayer membranes and recent results for
synthetic colloidal systems~\cite{Gompper2017,Gambin2006,
StebeCurvatureRodAssemblyPNAS2011,DesernoMuellerGuven2005}, we consider
diffusion limited interactions by particles that interact within a curved fluid
interface.  We investigate the case with and without hydrodynamic coupling on
the collective diffusivity and how this influences the time for particles to
come into near contact.

\begin{figure}[H] \centering
\includegraphics[width=1.0\columnwidth]
{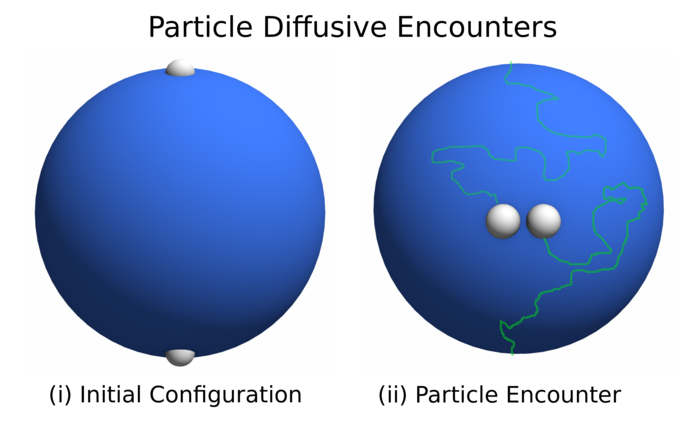}
\caption{Diffusive Encounters between Two Particles.  We show a schematic of
our study of the distribution of times for two diffusing particles to come into
contact.  The two particles start respectively at the north and south poles
(left).  In the cases with and without hydrodynamic coupling, we investigate
the distribution of times for when the particles come into diffusive contact
(right).} \label{fig:cluster_pair_meeting_times} \end{figure}

We consider two particles initially at antipodal locations on the sphere
(north and south pole) and the amount of time it takes for them to come into
contact with each other.  We consider the distribution for this meeting time in
the case of hydrodynamic coupling with mobility $M$ as in
Section~\ref{sec:autocorrelation_fluct_hydro} and in the case without
hydrodynamics with a local drag having mobility response $M = -\gamma
\mathcal{I}$.  We report these results with and without hydrodynamic
interactions in Figure~\ref{fig:cluster_pair_meeting_times}.  We use in our
studies the parameter values in Table~\ref{table:cluster}.

The $R$ gives the radius of the spherical fluid 
interface, the $R_{particle}$ gives the effective size 
of the particles used to specify the radius of separation
for considering contacts, $D$ is the particle diffusivity, 
and $\gamma_{slip}$ is the slip permitted between the 
particle and the local fluid.  The $L^{\pm} = \mu_m/2\mu_{\pm}$ 
gives the Saffman-Delbr\"{u}ck lengths between the surface fluid
viscosity $\mu_m$ and the surrounding bulk fluid viscosity $\mu_{-}$
inside and $\mu_{+}$ outside the spherical fluid interface.  The 
$k_B{T}$ gives the thermal energy.

%%% Table of parameters
\begin{table}[H] \small \centering \ra{1.1} 
\scalebox{0.9}{%
\begin{tabular}{|ll||ll|}
%\toprule
\hline \rowcolor{LightGrey} Parameter & Value & Parameter & Value\\
%\midrule
$R$ & 15.3 nm &$R_{particle}$ &1.5 nm\\ 
\hline $k_BT$ &2.48 amu nm$^2$ / ps &$D$ &3e-7 nm$^2$/ps\\ 
\hline $\gamma_{drag}^{-1}$ &$D/k_BT$ &$\gamma_{slip}$ &0.15 \\ 
\hline $\tau_D$ &$R^2/D$  &$\Delta t$ & 5e-6$\tau_D$ \\ 
\hline $L^+$ & 100 & & \\ 
\hline
\end{tabular}} 
\caption{Parameters used for the pair diffusivity study and
simulations.  We use SI units of atomic-mass-units (amu), nanometers
(nm), and picoseconds (ps).} 
\label{table:cluster} 
\end{table}

We find in both cases that the meeting times remain on average on the same
order of magnitude.  However, the hydrodynamic coupling introduces
significantly more variation in the meeting-time distribution producing a
long-tail, see Figure~\ref{fig:cluster_pair_meeting_times}.  For systems where
diffusive kinetics are important, we see the hydrodynamic coupling can
significantly influence the distribution of encounters between particles.

\begin{figure}[H] \centering
\includegraphics[width=1.0\columnwidth]
{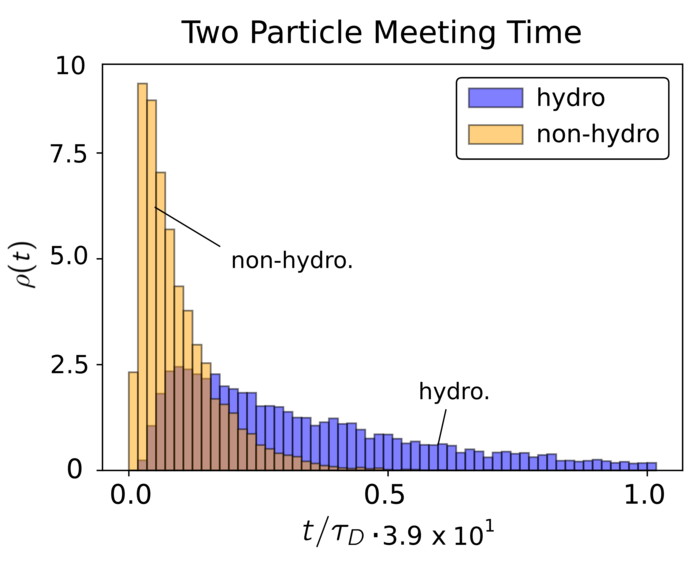}
\caption{Two Particle Pair-Meeting Time Distribution.  For two particles
starting at antipodal locations on the sphere, we show the distribution of
times for the two particles to come into contact at critical distance $r \leq
\tilde{r}_c = 3R_{particle}$.} \label{fig:cluster_pair_meeting_times}
\end{figure}

\subsection{Microscopic Swimmers and Mixing} \label{sec:micro_swimmer} We
investigate hydrodynamic transport and diffusive mixing associated with
swimmers at microscopic scales.  Behaviors both individually and collectively
can differ significantly from macroscopic
scales~\cite{LaugaPowersSwimmers2009,Purcell1977}.  Swimming in both Newtonian
and Non-Newtonian bulk fluids in three dimensional volumes have been
investigated in~\cite{LaugaPowersSwimmers2009,Purcell1977,Golestanian2004}.  We
investigate here the case of swimmers within two dimensional curved fluid
interfaces treated as a Newtonian fluid.  Our approaches also could be used to
study non-Newtonian fluids either by incorporating explicitly the
microstructures in the fluid using approaches of Section~\ref{sec:IB_curved}
and~\cite{AtzbergerShear2013}, or by extending the formulation of the
hydrodynamic equations to other constitutive laws in
Section~\ref{sec:conservation_laws}.

As a demonstration of the methods, we consider Golestanian
Swimmers~\cite{Golestanian2004} that consist of three beads that interact
through two oscillating harmonic bonds.  The harmonic bonds have time-dependent
rest-lengths with energy \begin{eqnarray} E_i(r) =\dfrac{1}{2}k[r -
(l+A\sin(\omega t +\phi_i))]^2, \end{eqnarray} where $i = 0,2$.  The bond
lengths are offset in time to have different phases for $\phi_0$ and $\phi_2$.
To impose excluded volume, both the beads of the microscopic swimmers and the
passive particles also interact through the Weeks-Chandler-Andersen (WCA)
potential~\cite{Weeks1971} \begin{align} \label{eqn_wca_potential}
\small{U_{\mbox{\tiny wca}}(r) = \begin{cases}
4\epsilon_{wca}\left[\parz{\dfrac{\sigma_{wca}}{r}}^{12} \right. &\\
\hspace{0.6cm}\left. -\parz{\dfrac{\sigma_{wca}}{r}}^6\right],
&r\leq2^{1/6}\sigma_{wca} \\ -\epsilon_{\mbox{\tiny wca}}, &
r>2^{1/6}\sigma_{wca}. \end{cases} } \end{align} The separation distance is
denoted by $r = \|\mb{X}_i - \mb{X}_j \|$ for two particles $\mb{X}_i$ and
$\mb{X}_j$.  The potential gives an effective particle steric radius of $r_s
=2^{1/6}\sigma_{wca}$.  We give parameters for our model in
Table~\ref{table:swimmer}.  We remark that this parameterization is for
illustrative purposes of the methods and to obtain more physically realizable
systems may require further adjustments.  

%%% Table of parameters for Swimmer
\begin{table}[H] \small \centering \ra{1.1} 
\scalebox{0.9}{%
\begin{tabular}{|ll||ll|}
%\toprule
\hline \rowcolor{LightGrey} Variables & Values & Variables & Values\\
%\midrule
$R$ & 15.3nm &$R_{particle}$ &1.5nm\\ 
\hline $k_BT$ &2.48 amu nm$^2$ / ps$^2$ &$D$ &9.4577e-1 nm$^2$ / ps\\
\hline $\tau_D$ &$R^2/D$  &$\Delta t$ & 5e-7$\tau_D$ \\ 
\hline $t$  &$5e5\Delta t$ &$\gamma_{slip}$ &2.6222e-3 ps/amu \\ 
\hline resI &100 &$\gamma$I &800\\ 
\hline $N$ & 3 particles &$k$ &$5e2 k_BT$ \\ 
\hline $\epsilon_{wca}$ &5$k_BT$  &$\sigma_{wca}$ &2$R_{particle}$ \\ 
\hline $\phi_0$ &0 &$\phi_2$ &$\pi/2$ \\ 
\hline $A$ &$R_{particle}$ &$l$ &2$R_{particle}$\\ \hline $\omega$ &5e-3 & & \\ 
\hline 
\end{tabular}} 
\caption{Parameter Values for the Swimmer
Simulations.  We use SI units of atomic-mass-units (amu), nanometers
(nm), and picoseconds (ps).} \label{table:swimmer} 
\end{table}

The phase differences in the swimming strokes are crucial to break symmetry in
time to have the possibility of a net forward motion.  For steady-state
hydrodynamics, a time-reversible motion would have no net displacement by the
Scallop Theorem~\cite{Purcell1977,Lauga2011,Ishimoto2012}.  We also emphasize
that without hydrodynamic coupling between the beads the swimmer would remain
stationary.  This is a consequence of the equal-and-opposite forces that act on
the beads and average out to zero over the periodic strokes. Without the
surrounding fluid, such forces can not move the center-of-mass of the swimmer.

\end{multicols} \begin{figure}[H] \centering
\includegraphics[width=1.0\columnwidth]{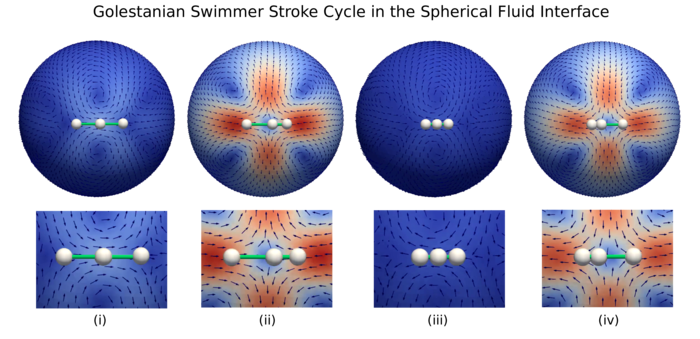}
\caption{Stroke Cycle and Hydrodynamic Flows of the Golestanian Microswimmer.
For the three bead Golestanian swimmer~\cite{Golestanian2004} immersed within a
spherical fluid interface, we show the configurations at each stage during the
swimming cycle and associated hydrodynamic flow for $L/R = 6.5$.  Different
flows are generated when the beads of the microswimmer are pushing or pulling
with respect to one another within the fluid.  The motions are by design
non-reversible in time and result in the microswimmer having net forward motion
to the right.  We show the average flow generated in the inset of
Figure~\ref{fig:swimmer_vel}.} \label{fig:swim_cycle}	\end{figure}
%\end{multicols}
\begin{multicols}{2} \noindent

\begin{figure}[H]
\centering
\includegraphics[width=1.0\columnwidth]
{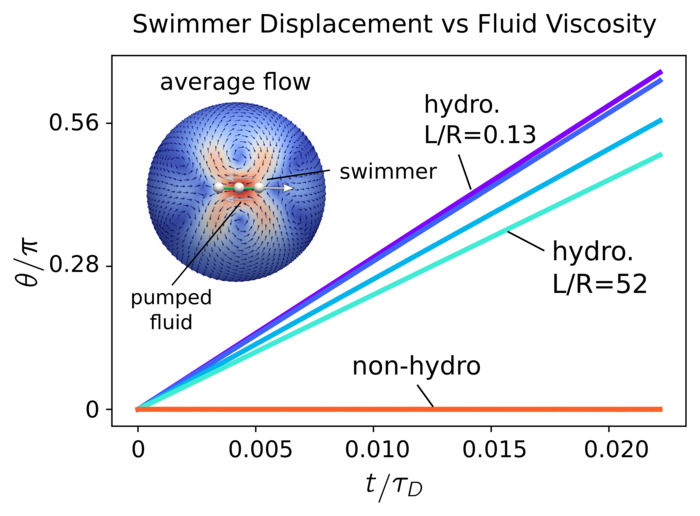}
\caption{Swimming Velocity on the Spherical Fluid Interface when varying the
Fluid Viscosities.  We show how the angular velocity and associated
displacement of microswimmers vary as the viscosity ratio $L/R$ increases. We
find the swimmer speed reduces over the viscosity range by $\sim 25\%$, see
Table~\ref{table:swimmer}.  We also show the results for the swimmer when
neglecting hydrodynamic coupling which as expected does not result in any
significant forward motion.  We show the time-averaged hydrodynamic flow
generated by the swimmer for $L/R = 6.5$ in the inset.  }
\label{fig:swimmer_vel}	\end{figure}

The swimmer strokes generate flows that on average pump the fluid.  We show the
associated stages of the stroke cycle captured by our methods in
Figure~\ref{fig:swim_cycle}.  The hydrodynamics is confined to a surface of
spherical topology which results in flows with vortices.  We show the average
flow that pumps the fluid in the inset of Figure~\ref{fig:swimmer_vel}.

We study first how the swimmer's speed depends on the viscosities of the two
dimensional interfacial fluid and surrounding bulk fluid.  We characterize the
viscosities by the ratio $L/R$ of the Saffman-Delbr\"{u}ck length $L$ to the
sphere radius $R$ as discussed in Section~\ref{sec:hydrodynamic_responses}.  We
can consider the swimmer's angular progression over the sphere from which the
swimming speed can be estimated, see Figure~\ref{fig:swimmer_vel}.  As the
viscosity ratio $L/R$ increases, the generated flows transition from being
relatively localized to enveloping most of the sphere.  We find as the
viscosity ratio increases from $L/R = 0.13$ to $L/R = 52$ that the swimming
speed drops by approximately $\sim 25\%$, see
Table~\ref{table:swimmerVelocity}.

We next investigate the collective drift-diffusion dynamics of $N_T$ passive
particles when subjected to mixing by multiple microscopic swimmers $N_S$.
Both the swimmers and particles are subjected to the hydrodynamic coupling and
thermal fluctuations using the approaches we introduced in
Section~\ref{sec:fluct_hydro_curved}.  We study as the number of microscopic
swimmers increases how the effective diffusivity of the passive tracer
particles is influenced.

We characterize the diffusivity by the Mean Squared Displacement (MSD)
\begin{eqnarray} MSD(t) = \left\langle \left\|X(t) - X(0)\right\|_g^2
\right\rangle.  \label{equ:swimmer_msd} \end{eqnarray} In the spherical
geometry the standard Euclidean distance is distorted by the spherical surface.
We use as our norm $\| \cdot \|_g$ the geodesic distance on the surface between
the starting point $X(0)$ and the final point $X(t)$.  Unlike bulk three
dimensional fluids, the distance between points remains bounded since the
surface is a compact manifold.  As a consequence, we have that eventually the
$MSD(t) \rightarrow m_0^2$ asymptotes to a limiting value.  This corresponds to
sampling the $X(t)$ from the stationary distribution over the surface.

\begin{figure}[H] \centering
\includegraphics[width=1.0\columnwidth]
{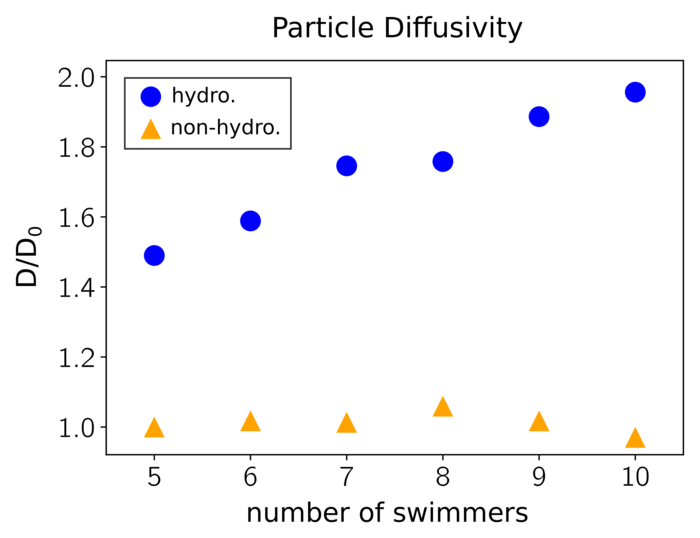}
\caption{Diffusivity verses Number of Swimmers.  We show how the enhanced
diffusivity depends on the number of microscopic swimmers.  We see that as the
number of swimmers increases the diffusivity is significantly enhanced from
$\sim 40\%$ to almost $\sim 100\%$ relative to the case without hydrodynamic
coupling where diffusivity remains constant.  The results are normalized by
$D_0 = 1.3703 nm^2/ps$ for the diffusivity in the case without
hydrodynamic coupling and $N_S = 5$ swimmers.  }
\label{fig:swimmer_diffusivity}	\end{figure}

To estimate in practice the diffusivity, we consider the $MSD$ over time-scales
$\tau$ with $\tau \leq 0.06\tau_D$ where $\tau_D$ is the time-scale to diffuse
the distance $R$.  We find the $MSD$ is approximately linear in this regime and
we characterize the passive particle motions by the diffusivity $D =
\partial{MSD(t)}/{\partial{t}}$, see Figure~\ref{fig:swimmer_diffusivity}.  In
practice, we estimate $D$ using the slope of the $MSD$ obtained from a
least-squares fit to the data in the range $0 \leq t \leq 0.06\tau_D$.

%%% Table of normalized velocities(heights in swimmer figure)
\begin{table}[H] \centering 
\begin{tabular}{|c|c||c|c|}
\rowcolor{LightGrey} \hline $L/R$  & $v/v_0$ (hydro) & $L/R$ & $v/v_0$ (hydro) \\ 
\hline 0.13 &1.167330  & 6.5 &1.000000 \\ 
\hline 0.65  &1.138532 & 52 &0.881074 \\ 
\hline
\end{tabular} 
\caption{Estimated swimmer velocity from the simulated swimmer
displacements with and without hydrodynamic coupling in
Figure~\ref{fig:swimmer_vel}.  Velocities normalized by $v_0 = 4.55\times10^3
nm/ps$ when $L/R = 6.5$.  The non-hydrodynamic swimmer simulations did
not exhibit any significant net motions $v/v_0 \sim 5.93 \times
10^{-4}nm/ps$.  We remark that this parameterization in
Table~\ref{table:swimmerVelocity} is for illustrative purposes of the methods
and to obtain more physically realizable systems may require further
adjustments.} 
\label{table:swimmerVelocity} 
\end{table}

We study how the diffusivity $D$ of $N_T = 80$ passive tracer particles are
enhanced by the action of $N_S$ swimmers.  Both the passive tracers and
swimmers undergo the drift-diffusive dynamics of
equation~\ref{equ_full_BD_model} with parameters in Table~\ref{table:swimmer}.
We remark there are some technical challenges in parameterizing consistently
for comparisons the non-hydrodynamic and hydrodynamic dynamics.  We choose to
do this by using the diagonal entry of our mobility tensor
$M_{\parallel\parallel}$ as the inverse drag $\gamma_{drag}^{-1}$ for the
non-hydrodynamic dynamics.  Given symmetries that can arise readily for small
numbers of swimmers, we consider cases with $N_S \in [5,10]$.  

Our initial studies reported here find the swimmers can result in significant
enhancement of the tracer diffusivity relative to the non-hydrodynamic case,
see Figure~\ref{fig:swimmer_diffusivity}.    In the case with no swimmers $N_S
= 0$, we have diffusivity with hydrodynamic coupling $D_{hy} = 1.70295
nm^2/ps$ and without hydrodynamic coupling $D_{nh} = 1.3916
nm^2/ps$.  This has the ratio $D_{hy}/D_{nh} = 1.2237$.  We see
already in the absence of any swimmers an effective enhancement of $\sim 20\%$
in the diffusivity of the particles from the hydrodynamic correlations.

\begin{figure}[H]
\centering
\includegraphics[width=1.0\columnwidth]
{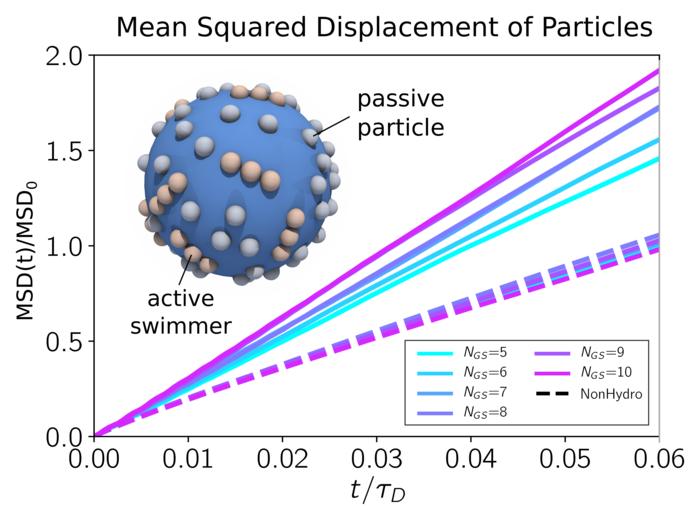}
\caption{Mean Squared Displacement (MSD) for Different Number of Swimmers.  We
show over time $t$ the $MSD(t)$ from equation~\ref{equ:swimmer_msd}.  We see
the hydrodynamic flows generated by the swimmers can significantly enhance the
$MSD$.  In contrast, in the absence of hydrodynamic coupling the $MSD$ of the
tracer particles changes relatively little.  The $MSD$ is normalized by $MSD_0
= MSD(t^*) = 0.1762 nm^2$ at time $t^* = 0.06$ for $N_S = 5$ without
hydrodynamics.  We show in the inset a typical configuration of the swimmers
and tracer particles for $N_T = 80$ and $N_S = 10$.  } \label{fig:swimmer_msd}
\end{figure}

As we introduce more microscopic swimmers $N_S$, their collective 
stroke cycles contribute hydrodynamic flows that mix 
the tracer particles in addition to the thermal fluctuations.
Relative to the non-hydrodynamic case, 
we find this manifests as an effective diffusivity that is 
further enhanced in the range of $\sim 40\%- 100\%$, see
Figure~\ref{fig:swimmer_diffusivity}.  The thermal fluctuations allow for the
tracer particles to diffuse between streamlines of the hydrodynamic flows that
are transiently generated by the swimmers.  The swimmers also can have their
own motions driven by the mutual flows and thermal fluctuations that serve both
to move their center-of-mass and to rotate their orientation.  This combination
of effects shows the interplay that can arise between hydrodynamically driven
drifts and thermal fluctuations.

The results illustrate some of the hydrodynamic and thermal 
effects that can be captured using our methods for possible 
further investigations of 
passive and active soft 
materials~\cite{MarchettiActiveSoftMaterials2013,Saintillan2018} 
or mechanics in cell biology~\cite{Boal2002,Mogilner2018}.
The introduced surface fluctuating hydrodynamics
methods can be used to model passive and active spatially extended
microstructures to capture both the hydrodynamic coupling and the correlated
thermal fluctuations within curved fluid interfaces.

\section{Conclusions} 
\label{sec:conclusions} 
We have introduced 
surface fluctuating hydrodynamics approaches for general 
investigations of 
the drift-diffusion
dynamics of particles and microstructures immersed within curved fluid
interfaces.  We introduced computational methods for simulations of
fluid-structure interactions and collective dynamics driven by active
forces and thermal fluctuations.  We studied the velocity autocorrelations
for surface fluctuating hydrodynamics in spherical geometries.  
We found for interfacial fluids there 
are different scalings that emerge for different physical regimes 
and depending on the interface geometry, surface viscosity, and
bulk viscosities.  We also showed how our methods can be used for 
modeling and simulating the collective drift-diffusion dynamics 
of both passive and active microstructures.  We obtained results
investigating the enhanced mixing of particles from 
active microswimmers.  The results show how the 
introduced surface fluctuating 
hydrodynamics approaches
can be used for investigating some of the rich phenomena that 
can arise in curved fluid interfaces. 

\section{Acknowledgments} The authors P.J.A, M.P. and D.A.R. acknowledge
support from research grants DOE Grant ASCR PHILMS DE-SC0019246 and NSF Grant
DMS - 1616353.  We also acknowledge UCSB Center for Scientific Computing NSF
MRSEC (DMR-1121053) and UCSB MRL NSF CNS-0960316.  P.J.A. would also like to
acknowledge a hardware grant from Nvidia.

\bibliographystyle{plain} \bibliography{paper_database}{}

\clearpage \newpage

\appendix

\section{Derivations for Fluctuating Hydrodynamics on Surfaces using Vector
Potentials $\Phi$} \label{appendix:derivation_fluct_hydro_phi} We discuss in
more detail for fluctuating hydrodynamics on the surface derivation of the
stochastic dynamics of the vector potential $\Phi$.  We express the dynamics in
terms of complex-valued coefficients $a_s$ of a spherical harmonics expansion
as \begin{eqnarray} \rho \frac{\partial a_s}{\partial{t}} = L_s a_s + \bar{c}_s
+ g_s + h_s.  \end{eqnarray} The force expansion terms $c_s^n$ are obtained
from the real-space force $f^n$ using equation~\ref{equ:Phi_fluct_hydro}.   We
obtain the stochastic driving fields $g_s$ and $h_s$ using a
fluctuation-dissipation approach for the discretized system in a manner similar
to our prior work~\cite{AtzbergerSIB2007,AtzbergerSELM2011}.  We generate
Gaussian driving terms with mean zero and covariance \begin{eqnarray}
\label{equ:g_fluc_corr} \left\langle g_s(t) g_{s'}(r) \right\rangle = -2 L_s
\mathcal{C}_{ss'} \delta(t - r).  \end{eqnarray} This requires determination of
the covariance $\mathcal{C}_{ss'}$ for the equilibrium fluctuations of the
modes $a_s$.

The Gibbs-Boltzmann distribution $\rho(\mb{v}^{\flat}) =
(1/Z)\exp\left(-E[\mb{v}^{\flat}]/k_B{T}\right)$ for the equilibrium
fluctuations depends on the kinetic energy of the fluid given by
\begin{eqnarray} E[\mb{v}^{\flat}] & = & \frac{\rho}{2} \int
\langle\mb{v}^{\flat},\mb{v}^{\flat}\rangle dA  \\ \nonumber & = &
\frac{\rho}{2} \int \langle -\star\mb{d} \Phi, -\star\mb{d} \Phi\rangle dA  \\
\nonumber & = & \frac{\rho}{2} \int -\langle \Phi, -\bs{\delta}\mb{d}
\Phi\rangle dA \\ \nonumber & = & \frac{\rho}{2} \sum_{s,s'} a_s a_{s'} \int
-\langle Y_s, -\bs{\delta}\mb{d}Y_{s'} \rangle dA \\ \nonumber & = &
-\frac{\rho}{2} \sum_{s} |a_s|^2 \lambda_{s} \|Y_s\|_2^2.  \end{eqnarray} The
$s = (\ell,m)$ where $\ell$ is the degree and $m$ the order.  We used here the
adjoint property of the co-differential $\bs{\delta}$ and the exterior
derivative $\mb{d}$~\cite{Abraham1988}.  The spherical harmonics modes $Y_s$
have $L^2$-norm given by $\|Y_s\|_2^2$ and are eigenfunctions of the
Laplace-Beltrami operator $-\bs{\delta}\mb{d} Y_s = \lambda_s Y_s$ with
eigenvalues $\lambda_{s} = -{\ell(\ell + 1)}/{R^2}$.  

The quadratic form of the energy yields that the equilibrium fluctuations of
the spherical harmonics coefficients are Gaussian with mean zero and covariance
\begin{eqnarray} \mathcal{C}_{ss'} = \left\langle a_s a_{s'} \right\rangle =
\rho^{-1} k_B{T} |\lambda_s|^{-1} \delta_{s',\overline{s}}.  \end{eqnarray} The
$\delta_{s',\overline{s}}$ denotes the Kronecker $\delta$-function where we use
notation $\overline{s} = (\ell,-m)$ to denote the conjugate mode index.  The
spherical harmonics coefficients are complex-valued and the field $\Phi$ must
be real-valued.  This requires for the coefficients $\overline{a_s} =
a_{\overline{s}}$.

We generate the stochastic driving terms using $g_s = \eta_s + i\xi_s$ with $i
= \sqrt{-1}$ and $\overline{g_s} = g_{\overline{s}}$ throughout. From the
conditions in equation~\ref{equ:g_fluc_corr}, we have for $m \neq 0$ that
\begin{eqnarray} \left \langle g_s g_{\overline{s}}\right\rangle  = \left
\langle g_s \overline{g_s}\right\rangle  = \eta_s^2 + \xi_s^2 =
-2L_s\mathcal{C}_{s,\overline{s}}. \\ \left \langle g_s g_s\right\rangle =
\left \langle \eta_s^2 - \xi_s^2 + 2i \eta_s\xi_s \right\rangle = 0.
\end{eqnarray} As a consequence, we must have that $\left \langle \eta_s^2
\right\rangle = \left \langle \xi_s^2 \right\rangle$ and $\left \langle \eta_s
\xi_s \right\rangle = 0$.  This requires that $\left \langle \eta_s^2
\right\rangle = \left \langle \xi_s^2 \right\rangle =
-L_s\mathcal{C}_{s,\overline{s}}$.  The case with $m = 0$ is special since the
mode is self-conjugate requiring \begin{eqnarray} \left \langle g_s
g_{\overline{s}}\right\rangle = \left \langle g_s g_s \right\rangle = \left
\langle g_s \overline{g_s}\right\rangle = \eta_s^2 + \xi_s^2 =
-2L_s\mathcal{C}_{s,\overline{s}} \\ \left \langle g_s g_s\right\rangle = \left
\langle \eta_s^2 - \xi_s^2 + 2i \eta_s\xi_s \right\rangle = -2L_s
\mathcal{C}_{s,\overline{s}}.  \end{eqnarray} As a consequence, we must have
that $\left \langle \xi_s^2 \right\rangle = 0$ and $\left \langle \eta_s^2
\right\rangle = -2L_s\mathcal{C}_{s,\overline{s}}$.  

Algorithmically, for modes $s = (\ell,m)$ these results correspond to
generating $g_s$ for $m > 0$ by computing each of the components $\eta_s$ and
$\xi_s$ as independent Gaussian random variates each having the covariance
$-L_s\mathcal{C}_{s\overline{s}}$, and for $m < 0$  setting $g_{\overline{s}} =
\overline{g_s}$.  The modes with $m = 0$ are special since they are
self-conjugate and we have $g_{\overline{s}} = \overline{g_s} = \eta_s$ with
covariance $-2L_s\mathcal{C}_{s\overline{s}}$.

\section{Derivation of Power-Laws for Fluctuating Hydrodynamics on a Sphere}
\label{appendix:autocor_v}

We find the autocorrelation of the fluid velocity on the spherical interface has
significantly different behaviors than bulk fluids.  For bulk Newtonian fluids
occupying a three dimensional volume the velocity autocorrelation function has
a well-known characteristic long-tail with scaling $\tau^{-3/2} $.  This is
supported by continuum theory~\cite{AtzbergerVelCorr2006,BedeauxMazur1974},
molecular simulations~\cite{AlderWainwright2D3DVelCor1970,Dorfman1972}, and
experimental evidence~\cite{PaulExpObsLongTail1981,Franosch2011}.  In contrast,
we find from equation~\ref{equ:fluct_hydro_fluid} and
equation~\ref{equ_Stokes_SPH_sol2_defA_ells} the interfacial fluid velocity
exhibits a few different intermediate power-law scalings and exponential decay
depending on the considered parameter and temporal regimes, see
Figure~\ref{fig:FluctHydroAutoCor}.  As we shall show, this arises both from
the spherical geometry and from the coupling between the two-dimensional
hydrodynamics and bulk surrounding three-dimensional fluid which introduce
additional time-scales.

The fluid velocity is obtained from $\Phi$ as $\mb{v} = (-\star \mb{d}
\Phi)^{\sharp} = v^{\theta} \partial_{\theta} + v^{\phi} \partial_{\phi}$ where
$v^{\theta} = \sum_s \frac{a_s}{\sqrt{g} R} \frac{\partial Y_s}{\partial \phi}$
and $v^{\phi} = -\sum_s \frac{a_s}{\sqrt{g} R} \frac{\partial Y_s}{\partial
\theta}$.  The velocity autocorrelation function associated with fluctuating
hydrodynamics on the sphere given in equation~\ref{equ:fluct_hydro_fluid} can
be expressed as \begin{eqnarray} \nonumber \left\langle v^{\theta}(0)
v^{\theta}(t) \right\rangle & = & \sum_s \sum_{s'} \frac{\left\langle a_s(0)
a_{s'}(t) \right\rangle }{|g| R^2} \frac{\partial Y_s}{\partial \phi}
\frac{\partial Y_{s'}}{\partial \phi}  \\ \nonumber & = & \sum_s \exp(tL_s)
\frac{k_B{T}}{|g|\rho \ell(\ell + 1)} \left|\frac{\partial Y_s}{\partial
\phi}\right|^2 \\ \nonumber \left\langle v^{\phi}(0) v^{\phi}(t) \right\rangle
& = & \sum_s \sum_{s'} \frac{\left\langle a_s(0) a_{s'}(t) \right\rangle }{|g|
R^2} \left|\frac{\partial Y_s}{\partial \phi}  \right|^2 \\ \nonumber & = &
\sum_s \exp(tL_s) \frac{k_B{T}}{|g|\rho \ell(\ell + 1)} \left|\frac{\partial
Y_s}{\partial \theta}\right|^2. \\ \label{equ:expand_autocor} \end{eqnarray}
For the spherical surface, we use that the metric $|g| = R^2$.  We also use
that $\langle a_s^2 \rangle = k_{B}T R^2/\rho \ell(\ell + 1)$ and $\langle
a_s(0) a_{s}(t) \rangle = \langle a_s^2 \rangle \exp(tL_s) $.

We consider the autocorrelation of the fluid at a point \begin{eqnarray}
\nonumber \left\langle \mb{v}(\mb{x},0)\cdot\mb{v}(\mb{x},t) \right\rangle =
\left\langle v^{\theta}(0)v^{\theta}(t) \right\rangle \| \partial_{\theta} \|^2
\\ + \left\langle v^{\phi}(0)v^{\phi}(t) \right\rangle \| \partial_{\phi} \|^2.
\end{eqnarray} Given the $\delta$-spatial correlation of the fluctuating
velocity field each of these series diverges in the limit $t\rightarrow 0$ when
the full infinite expansion is taken.  In practice in techniques such as the
Stochastic Immersed Boundary Methods~\cite{AtzbergerSIB2007} the fluctuating
velocity field is spatially averaged to model the dynamics of immersed
particles and microstructures.  This would result in fluid-structure
interactions over the surface only having effectively a responses to fluid
fluctuations above some critical length-scale (below some degree $\ell_*$)
which is related to the object's geometric size.  We can obtain in practice a
similar effect by working throughout with truncated series expansions with
$\ell \leq \ell_* = 50$~\cite{AtzbergerVelCorr2006}.

We always consider fluctuations at a point $\mb{x}_*$ on the equator for a
given spherical coordinate chart which by symmetry yields \begin{eqnarray}
\left\langle \mb{v}(\mb{x}_*,0)\cdot\mb{v}(\mb{x}_*,t) \right\rangle = 2 R^2
\left\langle v^{\phi}(0)v^{\phi}(t) \right\rangle \end{eqnarray} and
\begin{eqnarray} \left|\frac{\partial Y_s}{\partial \theta}\right|^2 = m^2
\left|Y_s\right|^2.  \end{eqnarray} We can express
equation~\ref{equ:expand_autocor} as \begin{eqnarray} \nonumber \left\langle
v^{\phi}(0) v^{\phi}(t) \right\rangle & = & \sum_\ell  \exp(tL_\ell)
\frac{k_B{T}}{|g|\rho \ell(\ell + 1)}  \cdot \\ \nonumber &\cdot& \sum_{|m|\leq
\ell}  m^2 \left|Y_s\right|^2.  \end{eqnarray} where $L_\ell$ is given by
equation~\ref{equ:L_s_same_mu}.

We approximate these sums asymptotically to estimate significant time-scales
governing different behaviors of the autocorrelation functions.  We make the
ansantz throughout that we can treat the term $\left|Y_s\right|^2 \sim C$.
This gives \begin{eqnarray} \nonumber \left\langle v^{\phi}(0) v^{\phi}(t)
\right\rangle & \approx & \sum_\ell  \exp(tL_\ell) \frac{k_B{T}}{|g|\rho
\ell(\ell + 1)}  \cdot \\ \nonumber &\cdot& C \sum_{|m|\leq \ell}  m^2.
\end{eqnarray} We use that $\sum_{|m|\leq \ell} m^2 = \ell(\ell + 1)(2\ell +
1)/3$.  This very conveniently cancels the term $\ell(\ell+1)$ in the
denominator that arose from the eigenvalues of the Laplace-Beltrami operator
discussed in Section~\ref{sec:hydrodynamic_responses}.  After some
rearrangement, we have \begin{eqnarray} \\ \nonumber \left\langle v^{\phi}(0)
v^{\phi}(t) \right\rangle & \approx & C \frac{k_B{T}}{|g|\rho} \sum_\ell
\left(\ell + \frac{1}{2}\right) \exp(tL_{\ell}).  \end{eqnarray} We have
absorbed also the additional prefactor constants into $C$.  The spherical
motion corresponding to rigid-body rotation in the bulk fluid corresponds to
the mode $\ell = 1$.  We see for $L_{\ell}$ only the second term (traction
stress term) persists in equation~\ref{equ:L_s_same_mu}.  This gives the decay
time-scale $\tau_r = {2RL}/{3\mu_m}$ for energy dissipation through the rigid
rotation of the entire spherical interface within the bulk fluid.  There are
two particularly interesting regimes.  The first corresponds to when $R/L \gg
1$ indicating localized hydrodynamics on the surface.  The second to when $R/L
\ll 1$ indicating the hydrodynamics is strongly coupled nearly over the entire
surface to give a response that is effectively a rigid-body rotation of the
sphere.  

We consider first the case when $R/L \gg 1$ and make the approximation
\begin{eqnarray} L_{\ell} \approx -\frac{\mu_m}{RL}\left[(\ell + \frac{1}{2}) +
\epsilon\right].  \end{eqnarray} The $\epsilon$ term includes the higher-order
terms.  We take $\ell\ll R/L$ so that the $\ell(\ell + 1)$ term does not play a
significant role.  We approximate the sum using integration to obtain
\begin{eqnarray} \nonumber \sum_{\ell} \left(\ell + \frac{1}{2}\right)
\exp\left(-\frac{t\mu_m}{RL}\left(\ell + \frac{1}{2}\right)\right)
\hspace{1.0cm}  \\ \nonumber \approx \int_1^{\ell_*}
\exp\left(-\frac{t\mu_m}{RL}\left(\ell + \frac{1}{2}\right)\right)\left(\ell +
\frac{1}{2}\right) d\ell \\ \nonumber = \int_{\frac{3}{2}}^{\ell_*}
e^{-t\alpha\tilde{\ell}} \tilde{\ell} d\tilde{\ell} =
\left[\frac{-1}{\alpha}e^{-\alpha t \tilde{\ell}} \tilde{\ell}
\right]_{3/2}^{\tilde{\ell}_*} \\ \nonumber
- \int_{\frac{3}{2}}^{\ell_*} \frac{-1}{\alpha t} e^{-\alpha t \tilde{\ell}}
  \tilde{\ell} d\tilde{\ell} \\ \nonumber = \left(\frac{3}{2\alpha t} +
\frac{1}{\alpha^2 t^2} \right)e^{-\frac{3}{2}\alpha t} \\ \approx
\frac{3}{2\alpha t} + \frac{1}{\alpha^2 t^2}. \hspace{1.45cm} \end{eqnarray} In
this notation, we set $\alpha = {\mu_m}/{RL}$.  We also make the assumption
that $t \ll \tau_r = {2RL}/{3\mu_m}$ so we can treat
$\exp\left({-\frac{3}{2}\alpha t}\right) \approx 1$.  In the case with $R/L \gg
1$ and $t \ll \tau_r = {2RL}/{3\mu_m}$, we have that \begin{eqnarray}
\left\langle v^{\phi}(0) v^{\phi}(t) \right\rangle \hspace{5cm} \nonumber \\
\nonumber \approx C \frac{k_B{T}}{|g|\rho} \sum_{\ell} \left(\ell +
\frac{1}{2}\right) \exp\left(-\frac{t\mu_m}{RL}(\ell + \frac{1}{2})\right)  \\
\approx \frac{k_B{T}}{|g|\rho} \frac{C}{\alpha^2 t^2}. \hspace{5.1cm}
\end{eqnarray} In the case approximating with $R/L \gg 1$ followed by
approximating with $\alpha = \mu_m/{RL} \gg 1$, we have that \begin{eqnarray}
\left\langle v^{\phi}(0) v^{\phi}(t) \right\rangle \hspace{5cm} \nonumber \\
\nonumber \approx C \frac{k_B{T}}{|g|\rho}\sum_{\ell}
\exp\left(-\frac{t\mu_m}{RL}(\ell + \frac{1}{2})\right) (\ell + \frac{1}{2})
\\ \approx \frac{k_B{T}}{|g|\rho} \frac{3C}{2\alpha t}.  \hspace{4.9cm}
\end{eqnarray} For the autocorrelation function, these results predict two
distinct regimes that will exhibit different power law scalings.  The first has
power law $t^{-2}$ and the second with power law scaling $t^{-1}$.  The
$t^{-1}$ power law is consistent with prior studies of pure two dimensional
fluid interfaces predicting similar results for the long-time tail and
divergence of the diffusion
coefficient~\cite{BedeauxMazur1974,AlderWainwright2D3DVelCor1970}.  Integrating
the velocity autocorrelation functions with appropriate truncations given
particle size one could obtain effective particle diffusivities within the
interface using the Green-Kubo relations~\cite{Kubo1957,Green1954,Reichl1998}.  

Given the coupling to the bulk surrounding fluid our fluctuating hydrodynamics
have additional time-scales mediating these effects.  It is interesting that
even though some of our parameter regimes exhibit a $t^{-1}$ decay this in fact
only persists for a finite amount of time and is eventually mitigated by our
coupling to the bulk solvent fluid.  Given that finite duration, our exhibited
$t^{-1}$ decay would result in finite logarithmic terms in the diffusivity,
which is consistent with Saffman-Delbr\"{u}ck theory which considers a similar
regime~\cite{SaffmanDelbruck1975}.  Our fluctuating hydrodynamics not only
capture the classical Saffman-Delbr\"{u}ck results but also extend this to
include geometric contributions from the spherical shape and other additional
time-scales in regimes where the interfacial hydrodynamics and coupling to the
bulk fluid could differ significantly.  Further extensions could also be made
to include the temporal dynamics of the bulk solvent fluid.

We next consider the regime with $R/L \ll 1$ and make the approximation
\begin{eqnarray} L_{\ell} \approx \frac{\mu_m}{R^2}\left[2 - \ell(\ell + 1) +
\epsilon\right].  \end{eqnarray} The $\epsilon$ term includes the higher-order
terms.  We see a key term is $\beta = \mu_m/R^2$.  In this regime we have
\begin{eqnarray} \nonumber \atzEquNumAndAdd \\ \nonumber \sum_{\ell}
\exp\left(\frac{t\mu_m}{R^2}(2 - \ell(\ell + 1) + \epsilon)\right) (\ell +
\frac{1}{2}) \\ \nonumber \approx \int_1^{\ell_*} \exp\left(t \beta(2 - \ell^2
- \ell) \right) (\ell + \frac{1}{2}) d\ell \\ \nonumber = \int_1^{\ell_*}
\exp\left(-t \beta(\ell + \frac{1}{2})^2 \right) (\ell + \frac{1}{2}) d\ell
\exp(t\beta\frac{9}{4}) \\ \nonumber = \left[ \exp\left(-t\beta(\ell +
\frac{1}{2})^2\right) \left( -\frac{1}{2t\beta} \right) \right]_{1}^{\ell_*}
\exp(t\beta \frac{9}{4}) \\ \nonumber = \exp(-t\beta \frac{9}{4}) \exp(t\beta
\frac{9}{4}) \left( \frac{1}{2t\beta} + \mbox{(small terms)} \right) \\
\nonumber \approx \frac{1}{2\beta t}. \hspace{6.8cm} \end{eqnarray} We obtained
these results using completion of the square in the exponent.

In the regime with $R/L \ll 1$, we see the velocity autocorrelation has
\begin{eqnarray} \nonumber \atzEquNumAndAdd \\ \left\langle v^{\phi}(0)
v^{\phi}(t) \right\rangle \hspace{6.5cm} \nonumber \\ \nonumber \approx
\frac{k_B{T}}{|g|\rho}  C\sum_{\ell} \exp\left(-\frac{t\mu_m}{R^2}(2 -
\ell(\ell + 1) + \epsilon)\right) (\ell + \frac{1}{2}) \\ \nonumber \approx
\frac{k_B{T}}{|g|\rho}  \frac{C}{2 \beta t}. \hspace{6.7cm} \end{eqnarray} This
predicts a power law decay with scaling $t^{-1}$.   

We see that the relaxation time-scale $\tau_r$ for some systems can be quite
large relative to the other time-scales.  In these regimes, we find something
interesting can occur where the velocity autocorrelation function plateaus.
This is predicted by our theory to occur in the regime when the time $t$
satisfies $\tau_a = {R^2}/{4\mu_m} \ll t  \ll \tau_r = {2RL}/{3\mu_m}$.  In
this regime, the correlations associated with the internal flow of the
hydrodynamics within the interfacial decays rapidly to zero.  However, the
rigid rotational motion of the entire fluid interface can still persist for
awhile until the rotation reaches its decay time-scale that dissipates this
motion.  This leads to the interesting plateaus seen in the velocity
autocorrelation function in Figure~\ref{fig:FluctHydroAutoCor}.  We also see
that for all non-zero parameter choices the autocorrelation function will
eventually exhibit an exponential decay when reaching time-scale $t \gg \tau_r
= {2RL}/{3\mu_m}$.  This is a consequence of the rigid-body rotational mode
$\ell = 1$ being the longest lived mode and eventually dissipating energy from
the interfacial fluid to the bulk surrounding fluid.  Our calculations show
that surface fluctuating hydrodynamics on quasi two dimensional fluid
interfaces can exhibit significantly different phenomena relative to their bulk
counter-parts in three dimensional space.

\section{Spherical Harmonics} \label{appendix:spherical_harmonics} We expand
functions $\Phi$ on the surface using the spherical harmonics \begin{eqnarray}
\label{equ:sphericalHarmonics_expansion} \Phi(\theta,\phi) =
\sum_{n=0}^{\infty} \sum_{m = -n}^{n} \hat{\Phi}_{n}^m Y_n^m(\theta,\phi),
\end{eqnarray} where \begin{eqnarray} \label{equ:sphericalHarmonics}
Y^m_n(\theta, \phi) & = & \sqrt{\frac{(2n+1)(n-m)!}{4\pi(n+m)!}}\cdot \\
\nonumber & \cdot & P^m_n\left(\cos(\phi)\right) \exp\left({im\theta}\right).
\end{eqnarray} The $m$ denotes the order and $n$ the degree for $n \ge 0$ and
$m \in \{-n, \dots, n\}$.  The $P^m_n$ denote the \textit{Associated Legendre
Polynomials}.  We denote by $\theta$ the azimuthal angle and by $\phi$ the
polar angle of the spherical coordinates~\cite{HanBookSphericalHarmonics2010}.
We work with real-valued functions and use that modes are self-conjugate in the
sense $Y^m_n = \overline{Y^{-m}_n}$.  

We can express the spherical harmonic modes as \begin{equation} Y^m_n(\theta,
\phi) = X^m_n(\theta, \phi)+ iZ^m_n(\theta, \phi).  \end{equation} The $X_n^m$
and $Z_n^m$ denote the real and imaginary parts.  We use this splitting in our
numerical methods to construct a purely real set of basis functions on the unit
sphere with maximum degree $N$ which consists of $(N+1)^2$ basis elements. For
the case $N=2$ we have the basis elements  \begin{eqnarray} \\ \nonumber
\tilde{Y}_1 = Y^0_0,\hspace{0.2cm} \tilde{Y}_2 = Z^1_1,\hspace{0.2cm}
\tilde{Y}_3 = Y^0_1,\hspace{0.2cm} \tilde{Y}_4 = X^1_1,\hspace{0.2cm} \\
\nonumber \tilde{Y}_5 = Z^2_2,\hspace{0.2cm} \tilde{Y}_6 = Z^1_2,\hspace{0.2cm}
\tilde{Y}_7 = Y^0_2,\hspace{0.2cm} \tilde{Y}_8 = X^1_2,\hspace{0.2cm} \\
\nonumber \tilde{Y}_9 = X^2_2.\hspace{0.2cm} \end{eqnarray} Similar conventions
are used for the basis for the other values of $N$.  We take final basis
elements $Y_i$ that are normalized as $Y_i = \tilde{Y}_i/\sqrt{\langle
\tilde{Y}_i, \tilde{Y}_i \rangle}$.

Derivatives are used within our finite expansions by evaluating analytic
formulas whenever possible for the spherical harmonics in order to try to
minimize approximation error~\cite{HanBookSphericalHarmonics2010}.
Approximation errors are incurred when sampling the values of expressions
involving these derivatives at the Lebedev nodes and when performing
quadratures~\cite{Lebedev1999}.  The derivative of the spherical harmonics in
the azimuthal coordinate $\theta$ is given by \begin{eqnarray} \\ \nonumber
\partial_\theta Y^m_n(\theta, \phi) &=& \partial_\theta
\sqrt{\frac{(2n+1)(n-m)!}{4\pi(n+m)!}} \cdot \\ \nonumber &\cdot&
P^m_n(\cos(\phi)) \exp\left({im\theta}\right) \\ \nonumber &=&
imY^m_n\left(\theta, \phi\right).  \end{eqnarray} We see this has the useful
feature that the derivative in $\theta$ of a spherical harmonic of degree $n$
is again a spherical harmonic of degree $n$.  As a consequence, we have in our
numerics that this derivative can be represented in our finite basis.  This
allows us to avoid additional $L^2$  projections allowing for computation of
the derivative in $\theta$ without incurring an approximation error.  The
derivative of the spherical harmonics in the polar angle $\phi$ is given by
\begin{eqnarray} \label{equ:derivPhi} \partial_\phi Y^m_n(\theta, \phi) & = & m
\cot(\phi)Y^m_n(\theta, \phi) \\ \nonumber & + & \sqrt{(n-m)(n+m+1)}\cdot \\
\nonumber & \cdot & \exp\left({-i\theta}\right) Y^{m+1}_n(\theta, \phi).
\end{eqnarray} We see that unlike derivatives in $\theta$ the derivative in
$\phi$ can not be represented in general in terms of a finite expansion of
spherical harmonics.  In our numerics, we use the expression in
equation~\ref{equ:derivPhi} for $\partial_\phi Y^m_n(\theta, \phi)$ when we
need to compute values at the Lebedev quadrature nodes.  These analytic results
provide a convenient way to compute derivatives of differential forms following
the approach discussed in our prior paper~\cite{AtzbergerGross2017}.  By using
these analytic expressions, we have that the subsequent hyperinterpolation of
the resulting expressions are where the approximation errors are primarily
incurred.  Throughout our discussions to simplify the notation we use the
convention that $Y^{m}_n = 0$ when $m \geq n+1$.  Further discussion of
spherical harmonics can be found~\cite{HanBookSphericalHarmonics2010}.  Further
discussions about how we use the spherical harmonics in our numerical
calculations of exterior calculus operators also can be found in our
papers~\cite{AtzbergerSoftMatter2016,AtzbergerGross2017}.

\section{Exterior Calculus: Coordinate Expressions}
\label{appendix:exterior_calc_coord}

We use approaches from exterior calculus 
to generalize operators used in continuum mechanics to
the manifold setting.  We give here more explicit 
expressions for these operators in terms of the local 
surface coordinates.  Since there is no
global non-singular coordinate system on the sphere, we ensure numerical
accuracy by switching between two coordinate charts.  In chart $A$ we have
coordinates $(\hat{\theta},\hat{\phi})$ with singularities at the north and
south poles.  In chart $B$ we have coordinates $(\tilde{\theta},\tilde{\phi})$
having singularities at the east and west poles.  To avoid issues with
singularities when seeking a value at a point $\mb{x}$, we evaluate expressions
within each chart in the regions with $\pi/4 \leq \phi \leq 3\pi/4$ and $\pi/4
\leq \tilde{\phi} \leq 3\pi/4$.  We give all expressions with generic polar
coordinates $(\theta,\phi)$ which we subsequently use in practice in our
numerical calculations by choosing the appropriate chart $A$ or chart $B$.
More details on our approach can also be found
in~\cite{AtzbergerSoftMatter2016}.  

The exterior derivative $\mb{d}$ for a $0$-form $f$ is given by 
$\mb{d}f = \partial f / \partial x_j \, \mb{x}_j$ and 
for a 1-form 
$\bs{\alpha} = \alpha^i d\mb{x}_i$ is 
$\mb{d}\bs{\alpha} = \partial \alpha^i/\partial x^j\, d\mb{x}_j \wedge d\mb{x}_i$.
The
$\bs{\delta}=-\star\mb{d}\star $ 
is the co-differential playing a role similar to the 
divergence on the surface~\cite{Abraham1988}.   The Hodge $\star$  
for a differential $k$-form $\bs{\beta}$ gives a complementary $n-k$-form 
$\star \bs{\beta}$ so that for any $k$-form $\bs{\alpha}$ we have
$\bs{\alpha}\wedge \star \bs{\beta} = \langle\alpha,\beta\rangle \bs{\omega}$
where $\bs{\omega}$ is the volume form~\cite{Abraham1988}.  This allows
us to generalize vector calculus operators such as the curl and divergence
to the surface by $\mbox{curl}_\mathcal{M}(\mb{v}^{\flat}) = -\star \mb{d} \mb{v}^{\flat}$
and $\mbox{div}_\mathcal{M}(\mb{v}^{\flat}) = \bs{\delta}\mb{v}^{\flat}$.

We now give coordinate expressions for these operations
in terms of the metric tensor and more specialized expressions in
the case of two dimensional manifolds (radial $2$-manifolds).
For radial $2$-manifolds, the exterior derivatives can be expressed
for a $0$-form $f$ and $1$-form
$\bs{\alpha}$ as \begin{eqnarray} \\ \nonumber \mb{d}f &=& (\partial_{\theta}f)
\mb{d}\theta + (\partial_{\phi} f) \mb{d}\phi = f_{\theta}\mb{d}\theta +
f_{\phi}\mb{d}\phi \\ \nonumber \mb{d}\alpha &=& (\partial_{\theta}
\alpha_{\phi} - \partial_{\phi} \alpha_{\theta}) \mb{d}\theta \wedge
\mb{d}\phi.  \end{eqnarray} 
The generalized curl in this setting for
$0$-form and $1$-form can be expressed as \begin{eqnarray} \nonumber -\star
\mb{d}f & = & \mbox{curl}_\mathcal{M}(f) \\ \nonumber &=&
\sqrt{|g|}\left(f_{\theta}g^{\theta \phi} + f_{ \phi} g^{\phi
\phi}\right)\mb{d}\theta \\ &-& \sqrt{|g|}(f_{\theta}g^{\theta \theta} +
f_{\phi}g^{\phi \theta})\mb{d}\phi \\ -\star \mb{d} \alpha & = &
\mbox{curl}_\mathcal{M}(\bs{\alpha}) = \frac{\partial_{\phi} \alpha_{\theta} -
\partial_{\theta} \alpha_{\phi} }{\sqrt{|g|}}.  \end{eqnarray} 
In this notation
we have taken the conventions that $f_j = \partial_{x^j} f$ and $\alpha_j$ such
that $\bs{\alpha} = \alpha_j \mb{d}{x}^j$ where $j \in \{\theta,\phi\}$.  The
isomorphisms $\sharp$ and $\flat$ between vectors and co-vectors can be
expressed explicitly as \begin{eqnarray} \mb{v}^{\flat} &=&
(v^{\theta}\bs{\sigma}_{\theta} + v^{\phi}\bs{\sigma}_{\phi})^{\flat}\\
\nonumber &=& v^{\theta}g_{\theta \theta}\mb{d}\theta + v^{\theta}g_{\theta
\phi}\mb{d}\phi + v^{\phi}g_{\phi \theta}\mb{d}\theta + v^{\phi}g_{\phi
\phi}\mb{d}\phi \\ \nonumber &=& (v^{\theta}g_{\theta \theta} + v^{\phi}g_{\phi
\theta})\mb{d}\theta + (v^{\theta}g_{\theta \phi} + v^{\phi}g_{\phi
\phi})\mb{d}\phi \\ \nonumber \\
(\bs{\alpha})^{\sharp} &=& (\alpha_{\theta}\mb{d}\theta +
\alpha_{\phi}\mb{d}\phi)^{\sharp} \\ \nonumber &=& \alpha_{\theta}g^{\theta
\theta}\bs{\sigma}_{\theta} + \alpha_{\theta}g^{\theta \phi}\bs{\sigma}_{\phi}
+ \alpha_{\phi}g^{\phi \theta}\bs{\sigma}_{\theta} + \alpha_{\phi}g^{\phi
\phi}\bs{\sigma}_{\phi} \\ \nonumber &=& (\alpha_{\theta}g^{\theta \theta} +
\alpha_{\phi}g^{\phi \theta})\bs{\sigma}_{\theta} + (\alpha_{\theta}g^{\theta
\phi} + \alpha_{\phi}g^{\phi \phi})\bs{\sigma}_{\phi} \end{eqnarray} We use the
notational conventions here that for the embedding map $\bs{\sigma}$ for
spherical coordinates in $\mathbb{R}^3$ we have $\bs{\sigma}_\theta =
\partial_{\theta}$ and $\bs{\sigma}_\phi = \partial_{\phi}$.  Combining the
above equations we can express the generalized curl as \begin{eqnarray}
\label{equ_gen_curl_0form} \\ \nonumber (-\star \mb{d}f)^{\sharp} &=&
([\sqrt{|g|}(f_{\theta}g^{\theta \phi} + f_{ \phi} g^{\phi \phi})]g^{\theta
\theta} \\ \nonumber & + & [- \sqrt{|g|}(f_{\theta}g^{\theta \theta} +
f_{\phi}g^{\phi \theta})]g^{\phi \theta})\bs{\sigma}_{\theta}\\ \nonumber &+&
([\sqrt{|g|}(f_{\theta}g^{\theta \phi} + f_{ \phi} g^{\phi \phi})]g^{\theta
\phi} \\ \nonumber & + & [- \sqrt{|g|}(f_{\theta}g^{\theta \theta} +
f_{\phi}g^{\phi \theta})]g^{\phi \phi})\bs{\sigma}_{\phi} \\ \nonumber &=&
\frac{f_{\phi}}{\sqrt{|g|}} \bs{\sigma}_{\theta} -
\frac{f_{\theta}}{\sqrt{|g|}} \bs{\sigma}_{\phi} \\ \nonumber \\
\label{equ_gen_curl_1form} \\ \nonumber -\star \mb{d}\mb{v}^{\flat} &=&
-\frac{\partial_{\phi}(v^{\theta}g_{\theta \theta} + v^{\phi}g_{\phi
\theta})}{\sqrt{|g|}} \\ \nonumber & + & \frac{\partial_{\theta}
(v^{\theta}g_{\theta \phi} + v^{\phi}g_{\phi \phi}) }{\sqrt{|g|}}.
\end{eqnarray} The scalar Laplace-Beltrami operator $\Delta_{LB} = -\bs{\delta}
\mb{d}$ that acts on $0$-forms can be expressed in coordinates as
\begin{equation} \Delta_{LB} = -\bs{\delta} \mb{d} =
\frac{1}{\sqrt{|g|}}\partial_i \left(g^{ij}\sqrt{|g|} \partial_j \right).
\end{equation} The $g_{ij}$ denotes the metric tensor, $g^{ij}$ the inverse
metric tensor, and $|g|$ the determinant of the metric tensor.

The velocity field of the hydrodynamic flows $\mb{v}$ is
recovered from the vector potential $\Phi$ as 
$\mb{v}^{\flat} = \mbox{curl}_\mathcal{M}(\Phi) = -\star \mb{d}
\Phi$.  The velocity field is obtained from $\mb{v} = \mb{v}^{\sharp} = \left(-\star
\mb{d} \Phi\right)^{\sharp}$ using equation~\ref{equ_gen_curl_0form}.
Similarly, from the force density $\mb{b}$ acting on the fluid, we obtain the
data $\mbox{curl}_\mathcal{M}(\mb{b}^{\flat}) = -\star\mb{d}\mb{b}^{\flat}$ 
for the vector potential formulation of the
hydrodynamics using equation~\ref{equ_gen_curl_1form}.  
Additional details and
discussions of these operators can be found in our related
papers~\cite{AtzbergerSoftMatter2016,AtzbergerGross2017} and
in~\cite{Pressley2001,Abraham1988, SpivakDiffGeo1999}.

\end{multicols}

\end{document}